\documentclass[amsmath, amssymb,10pt,aps,prx,twocolumn,notitlepage,showpacs,superscriptaddress]{revtex4-1}

\usepackage{graphics}
\usepackage{hyperref}
\usepackage{epsfig}
\usepackage{color}

\usepackage[normalem]{ulem}
\usepackage{multirow}
\usepackage{bm}

\begin{document}

\bibliographystyle{naturemag}

\title{Giant topological longitudinal circular photo-galvanic effect \\ in the chiral multifold semimetal CoSi}

\author{Zhuoliang Ni}
\affiliation{Department of Physics and Astronomy, University of Pennsylvania, Philadelphia, Pennsylvania 19104, USA}
\author{K. Wang}
\affiliation{Maryland Quantum Materials Center, Department of Physics, University of Maryland, College Park, MD 20742, USA.}
\author{Y. Zhang}
\affiliation{Max-Planck-Institut fur Chemische Physik fester Stoffe, 01187 Dresden, Germany}
\affiliation{Department of Physics, Massachusetts Institute of Technology, Cambridge, Massachusetts 02139, USA}
\author{O. Pozo}
\affiliation{Instituto de Ciencia de Materiales de Madrid, CSIC, Cantoblanco, Madrid, 28049, Spain}
\author{B. Xu}
\affiliation{Department of Physics and Fribourg Center for Nanomaterials, University of Fribourg, Chemin du Mus\'{e}e 3, CH-1700 Fribourg, Switzerland}
\author{X. Han}
\affiliation{Department of Physics and Astronomy, University of Pennsylvania, Philadelphia, Pennsylvania 19104, USA}
\author{K. Manna}
\affiliation{Max-Planck-Institut fur Chemische Physik fester Stoffe, 01187 Dresden, Germany}
\author{J. Paglione}
\affiliation{Maryland Quantum Materials Center, Department of Physics, University of Maryland, College Park, MD 20742, USA.}
\author{C. Felser}
\affiliation{Max-Planck-Institut fur Chemische Physik fester Stoffe, 01187 Dresden, Germany}
\author{A. G. Grushin}
\affiliation{Univ. Grenoble Alpes, CNRS, Grenoble INP, Institut N\'eel, 38000 Grenoble, France}
\author{F. de Juan}
\affiliation{Donostia International Physics Center, P. Manuel de Lardizabal 4, 20018 Donostia-San Sebastian, Spain}
\affiliation{IKERBASQUE, Basque Foundation for Science, Maria Diaz de Haro 3, 48013 Bilbao, Spain}
\author{E. J. Mele}
\affiliation{Department of Physics and Astronomy, University of Pennsylvania, Philadelphia, Pennsylvania 19104, USA}
\author{Liang Wu}
\email{liangwu@sas.upenn.edu}
\affiliation{Department of Physics and Astronomy, University of Pennsylvania, Philadelphia, Pennsylvania 19104, USA}

 \date{\today}

\begin{abstract}
The absence of mirror symmetry, or chirality, is behind striking natural phenomena found in systems as diverse as DNA and crystalline solids. A remarkable example occurs when chiral semimetals with topologically protected band degeneracies are illuminated with circularly polarized light. Under the right conditions, the part of the generated photocurrent that switches sign upon reversal of the light's polarization, known as the circular photogalvanic effect, is predicted to depend only on fundamental constants. The conditions to observe quantization are non-universal, and depend on material parameters and the incident frequency. In this work, we perform terahertz emission spectroscopy with tunable photon energy from 0.2 eV - 1.1 eV in the chiral topological semimetal CoSi. We identify a large longitudinal photocurrent peaked at 0.4 eV reaching $\sim$ 550 $\mu A/V^{2}$, which is much larger than the photocurrent in any chiral crystal reported in the literature. Using first-principles calculations we establish that the peak originates from topological band crossings, reaching 3.3$\pm$0.3 in units of the quantization constant. Our calculations indicate that the quantized CPGE is within reach in CoSi upon doping and increase of the hot-carrier lifetime. The large photo-conductivity suggests that topological semimetals could potentially be used as novel mid-infrared detectors.
\end{abstract}

\maketitle

Circular photogalvanic effect (CPGE) exists only in gyrotropic crystals \cite{sturmanbook1992, ganichevJPCM2003}. Its transverse component, where the current flows perpendicular to light propagation direction, is by far the most commonly observed. It has been recently measured in transition metal dichalcogenides \cite{yuanNatureNano2014}, topological insulators~\cite{mciverNatureNano2012, olbrichPRL2014, okadaPRB2016} and Weyl semimetals \cite{maNatPhys2017, siricaPRL2019, gaoNatComm2020}. In contrast, the longitudinal CPGE, where current flows parallel to light propagation direction, remains more elusive since its discovery in tellurium in 1979 \cite{asnin1979circular}. 

\begin{figure*}
\centering
\includegraphics[width=0.7\textwidth]{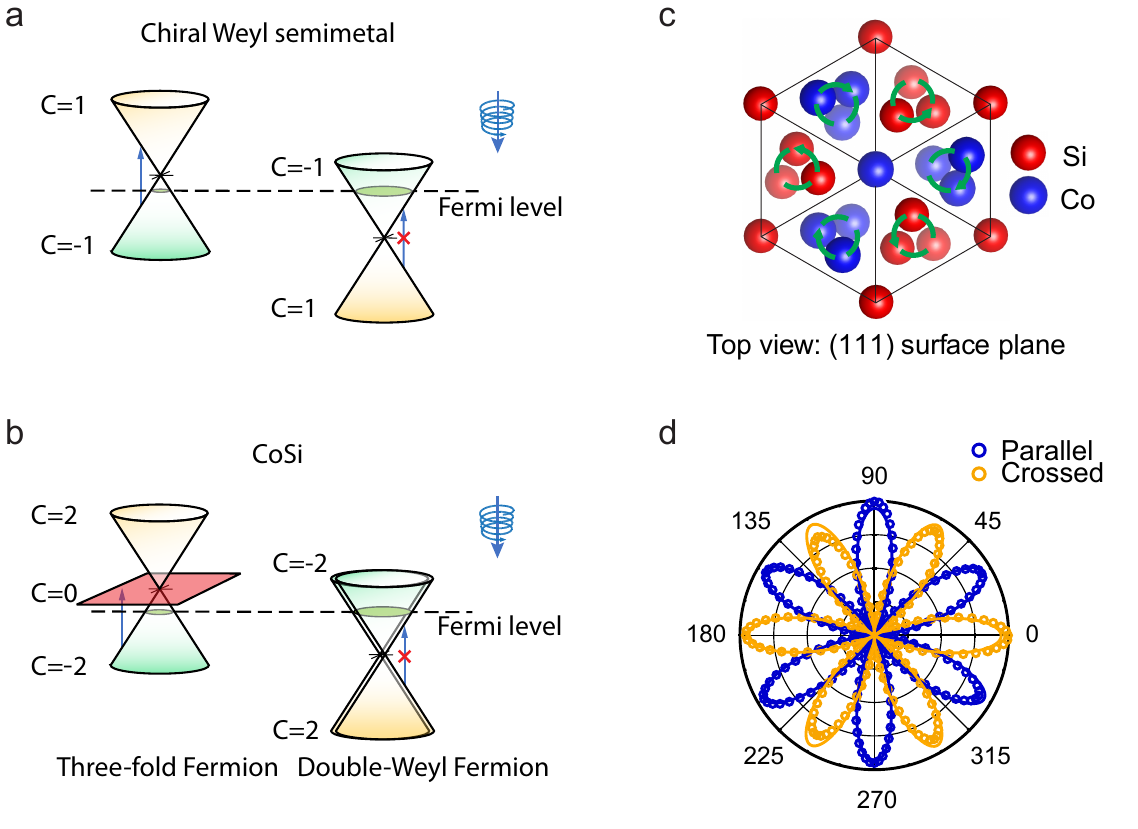}
\caption{\textbf{Schematic diagrams of the crystal and band structure of topological chiral semimetals, and the second haromonic generation on CoSi.} 
\textbf{a}, (Top) A chiral Weyl semimetal has two Weyl nodes with opposite monopole charges $\pm1$  at different energies. (Bottom) Without spin-orbit coupling CoSi features a threefold node and a double Weyl node located at different energies and momenta. The three bands in the threefold fermions have Chern numbers +2, 0 and -2 respectively. When spin degeneracy is accounted for, the total charges at the two modes are $\pm$4. When circularly polarized light is incident on the sample, excitations around the right Weyl (top) or the double Weyl fermion (bottom) are Pauli blocked, but excitations around the left Weyl (top) or the threefold fermions (bottom) are allowed, generating a CPGE.
\textbf{b}, Schematic top view of CoSi (111) surface. The different transparency of Si and Co atoms indicates the different depth of each atom from the top plane. The green anticlockwise/clockwise circles indicate the chirality of the Co/Si atoms.
\textbf{c}, Second harmonic generation signal generated from CoSi (111) natural facet under normal incident laser pulses with a photon energy of 1.55 eV. Dots are experimental data from parallel and crossed configurations between incident and detecting linear polarization. Solid lines are the best fit.
}
\label{Fig1}
\end{figure*}

In chiral topological semimetals, the longitudinal CPGE is particularly remarkable because it was recently predicted to be quantized~\cite{dejuanNatComm2017,changPRL2017,flickerPRB2018, deJuan:2020jm}. These materials feature protected nodal crossings near the Fermi level, and because all mirror symmetries are broken, nodes with opposite chirality generically appear at different energies \cite{huangPNAS2016} (see Fig.\ref{Fig1}\textbf{a}), in contrast to mirror-symmetric Weyl metals, like TaAs with nodes at the same energy~\cite{WengPRX2015, HuangNatComm2015}. The existence of these  nodes is protected  by an integer topological charge $C$, which quantizes the longitudinal CPGE trace to $C\beta_0$ where $\beta_0=\pi e^3/h^2$~\cite{dejuanNatComm2017}.  Chiral Weyl metals, where $C=\pm 1$ {(Fig. \ref{Fig1}\textbf{a} top)}\cite{huangPNAS2016}, are elusive.  Nevertheless, separated topological nodes with degeneracies larger than two, known as multifold fermions, are demonstrated to exist in chiral crystals such as CoSi, RhSi and AlPt (with $C=\pm 4$) \cite{bradlynScience2016,tangPRL2017,changPRL2017, raoNature2019, sanchezPRL2019,takanePRL2019,schroterNatPhys2019} (Fig. \ref{Fig1}\textbf{a} bottom). Furthermore, the presence of cubic symmetry in these materials makes transverse CPGE vanishing and longitudinal CPGE isotropic with only one non-zero independent term, $\beta_{xx}$, so that averaging over the three directions is not needed to measure the tensor trace $\beta$ ($\beta$=3$\beta_{xx}$).

Several challenges remain to observe the quantized CPGE in chiral semimetals. In this family of materials, the presence of spin-orbit coupling leads to a splitting of the nodes, which can still display quantization but in a reduced frequency range determined by the strength of the spin-orbit coupling, for example as happens in RhSi \cite{ changPRL2017,flickerPRB2018,reesarxiv2019, deJuan:2020jm, NiarXiv2020}. {Inter-band excitations between the spin-orbit split bands contribute to the non-quantized CPGE, a non-negligible effect in RhSi \cite{deJuan:2020jm}. Therefore a small spin-orbit splitting is  advantageous to observe the quantized CPGE in multifold materials~\cite{dejuanNatComm2017,flickerPRB2018}.} In this work, we measure the CPGE in CoSi as its spin-orbit coupling is much smaller than in RhSi \cite{tangPRL2017,changPRL2017}.

We measure the CPGE by detecting radiated terahertz (THz) pulses emitted from the illuminated regions, a method with several advantages compared to DC current measurements ~\cite{braunNatComm2016,sotomePNAS2019, reesarxiv2019, siricaPRL2019, gaoNatComm2020, NiarXiv2020}. Firstly, {detecting CPGE in a contact-less way avoids  contact misalignment as we measure second harmonic generation to align the crystal axis.} Secondly, the emitted THz pulse originates from a local illuminated region, and therefore thermal current and the non-local diffusion of photo-excited carriers to the contacts, typical of a DC current measurement, are not concerns~\cite{songPRB2014}. Thirdly, in the process of THz emission, photo-carriers move at the band velocity and then recombine, which creates a time-dependent photo-current within the penetration depth. This time-dependent current radiates a THz pulse into free space, which in the far field is 
 related to the first time-derivative of the surface current and is directly related with the quantized CPGE as  the quantized quantity is the rate of the inject current, $dj/dt=iC\beta_0$, instead of the current $j$ itself \cite{shan2004terahertz}. Note that optical rectification is generally at least two orders of magnitude smaller than the photocurrent in the resonant regime \cite{nastosPRB2006, sotomePNAS2019}. Therefore, we neglect the optical rectification effect in CoSi.

In this work, we have developed the capability to measure THz emission in the mid-infrared regime (0.20 eV - 0.48 eV) for the first time. We use it to measure CPGE in CoSi across a broad range of 0.2 eV - 1.1 eV. We identify a large longitudinal CPGE peaked at 0.4 eV reaching $\sim$ 550 $\mu A/V^{2}$. Comparing our measurements to  first-principles calculations, we establish that the peak originates from topological band crossings, reaching 3.3$\pm$0.3 in units of the quantization constant under the assumption of a constant hot-carrier lifetime.  We develop a $k \cdot p$ model, including quadratic corrections to the dispersion of the nodal bands, and show that the location of the chemical potential can conspire to create a more complex frequency profile than it has been anticipated even in the spinless model. Our calculations also lay out the conditions to observe the quantized CPGE  in CoSi in future experiments.
\\

\textbf{Results}

\begin{figure*}
\centering
\includegraphics[width=\textwidth]{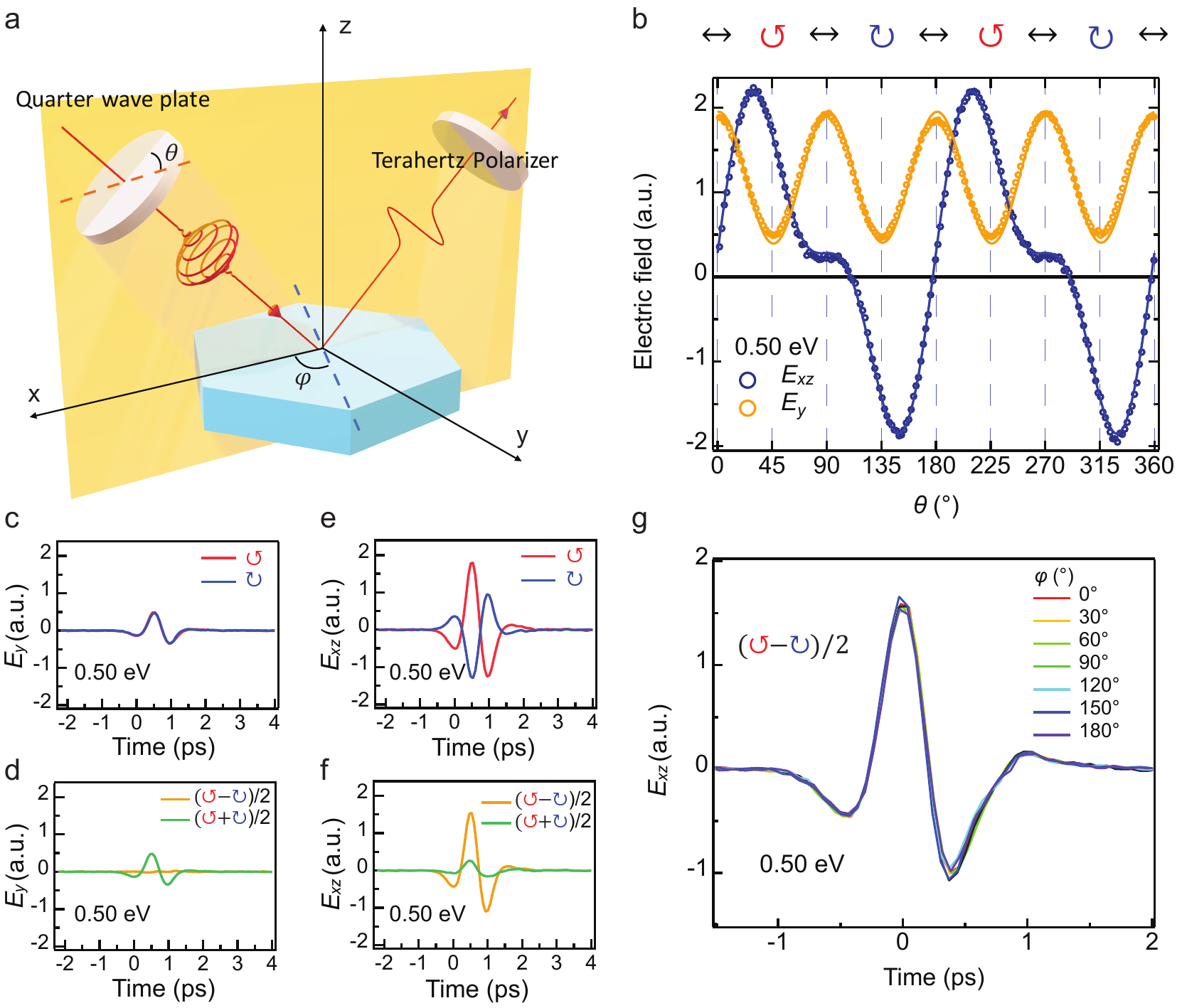}
\caption{\textbf{Schematic depiction of the CPGE experiment and experimental data on CoSi.} 
\textbf{a}, Schematic diagram of experimental setup. 
\textbf{b}, A typical set of $xz$  (in-plane) and $y$ (out-of-plane) components of the peak of the emitted THz time trace as a function of the angle of the quarter-wave plate under light pulses centered at 0.50 eV. The open circles are experimental data and the lines are the best fit constrained by the crystal symmetry of CoSi. 
\textbf{c}-\textbf{f}, A typical set of THz time traces of the out-of-plane  (\textbf{c}) and in-plane (\textbf{e}) components under the left-handed and right-handed incident pulses at 0.50 eV. Curves in \textbf{d} and \textbf{f} describe the extracted contribution for CPGE ($(E_{\circlearrowleft}-E_{\circlearrowright})/2$, orange) and LPGE ($(E_{\circlearrowleft}-E_{\circlearrowright})/2$, green). \textbf{g}, Nearly identical CPGE THz time traces at different sample azimuth angles $\phi$ at the incident photon energy of 0.50 eV. }
\label{Fig2}
\end{figure*}

\textbf{Sample and Second Harmonic Generation.} The chiral crystal structure of CoSi seen from the (111) direction is depicted in Fig.~\ref{Fig1}\textbf{b}. As a first step, we pick up large homogeneous (111) natural facets by scanning second-harmonic generation (SHG) measurement~\cite{wuNatPhys2017}. To stimulate SHG, we focused  light pulses centered at 800 nm under near-normal incidence to a 10-$\mu$m diameter spot on the sample and the second haromic signal centered at 400 nm is measured. As shown in Fig.~\ref{Fig1}\textbf{c}, polar patterns of SHG are found as a function of the direction of the linear polarization of the incident light in the co-rotating parallel-polarizer (orange) and crossed-polarizer (blue) configurations. These patterns agree well with a fit with only one non-zero parameter based on the point group symmetry (see Methods). \\

\textbf{Longitudinal CPGE in CoSi.} Fig.~\ref{Fig2}\textbf{a} shows schematically the measurement of the longitudinal CPGE. When circularly polarized light is incident on the sample, a current flows  along the light propagation direction inside of the material. Under normal incidence, the current flows perpendicular to the surface, which prevents THz emission into free space from CPGE in the bulk~\cite{reesarxiv2019, shan2004terahertz}. {THz emission does originate from the surface current density under oblique incidence~\cite{shan2004terahertz}. See SI section \textbf{I} for more details.} Therefore, in order to emit THz radiation into free space, we utilize off-normal incidence at 45 degrees as shown in  Fig.~\ref{Fig2}\textbf{a}. An achromatic quarter-wave plate is used to control the polarization of the incident light. Terahertz wave components in \textit{xz} and \textit{y} direction are detected by using a THz polarizer before the ZnTe detector.  Fig.~\ref{Fig2}\textbf{b,e} shows the reversal of the polarity of the time trace of emitted THz electric field under left and right circularly-polarized light at the incident photon energy of 0.50 eV, which indicates the change of the direction of the photo-current under opposite helicity of circularly polarized incident light (i.e. the CPGE).

To confirm that the CPGE we observe is a longitudinal photocurrent, we studied the polarization dependence of the CPGE by rotating both the achromatic quarter-waveplate and the samples. The experimental geometry is shown in Fig.~\ref{Fig2}\textbf{a}, and we detect the emitted THz components in the incident plane, $E_{xz}(t)$, and perpendicular to the plane, $E_y(t)$. In Fig.~\ref{Fig2}\textbf{b} (orange) we show the peak value of the emitted THz field $E_y(t)$ at the incident photon energy of 0.50 eV as a function of the angle of the quarter-wave plate. The THz field under left and right circularly polarized light has the same direction and magnitude, which indicates no CPGE.  The almost identical time traces  of $E_y(t)$ with opposite circular polarizations are shown in Fig.~\ref{Fig2}\textbf{c}. The CPGE component $(E_{\circlearrowleft}(t)-E_{\circlearrowright}(t))/2$ is zero within the detection sensitivity, as shown in Fig.~\ref{Fig2}\textbf{d}.  $(E_{\circlearrowleft}(t)+E_{\circlearrowright}(t))/2$ is  the linear photogalvanic effect (LPGE) component under circularly-polarized light (see supplementary information (SI) section \textbf{I} B for details).

In contrast, the  in-plane THz  field $E_{xz}(t)$ shows completely different polarization dependence as shown in Fig.~\ref{Fig2}\textbf{b} (blue). When the helicity of the circularly polarized light is reversed, the direction of the peak THz field in $E_{xz}(t)$ changes, and the waveform  is shown in Fig.~ \ref{Fig2}\textbf{e}. They are not simply the same curve with opposite signs because of a sizable LPGE contribution.   Nevertheless,  $(E_{\circlearrowleft}-E_{\circlearrowright})/2$  is not zero in $E_{xz}(t)$ and relatively large compared to $(E_{\circlearrowleft}+E_{\circlearrowright})/2$, as shown in  Fig.\ref{Fig2}\textbf{f}. The observation of a non-zero CPGE only in the incident $xz$ plane is consistent with the longitudinal CPGE, where the current flows along the  wave vector direction inside CoSi. This longitudinal CPGE is unchanged as we rotate the sample due to the cubic symmetry constraints, as shown for 0.50 eV incident photon energy in Fig.\ref{Fig2}\textbf{g}. We also observed similar angle dependence at other photon energies. See SI section \textbf{I} B for more details.

 To quantify the longitudinal CPGE, we performed a symmetry analysis by fitting the angle dependence of the quarter-wave plate on $E_{xz}$ and $E_y$. The solid lines in Fig.~\ref{Fig2}\textbf{b} are the best fit to the functions determined by the crystal symmetry, $E_y(\theta)=B_1\sin(4\theta)+C_1\cos(4\theta)+D_1$ and $E_{xz}(\theta)=A\sin(2\theta)+B_2\sin(4\theta)+C_2\cos(4\theta)+D_2$, where the coefficients $A$,$B_1$,$B_2$,$C_1$,$C_2$,$D_1$,$D_2$ are determined by the CPGE and LPGE conductivity. (see SI section \textbf{I} B  for details).
Both curves are fitted simultaneously with the same fit weights.  The $\sin(2\theta)$ term describes the CPGE while $\sin(4\theta)$,  $\cos(4\theta)$ and the constant terms describe the LPGE. The symmetry analysis shows that the out-of-plane component $E_y$ does not contribute to the CPGE, while the in-plane component $E_{xz}$ dominates the CPGE. \\

\begin{figure*}
\centering
\includegraphics[width=\textwidth]{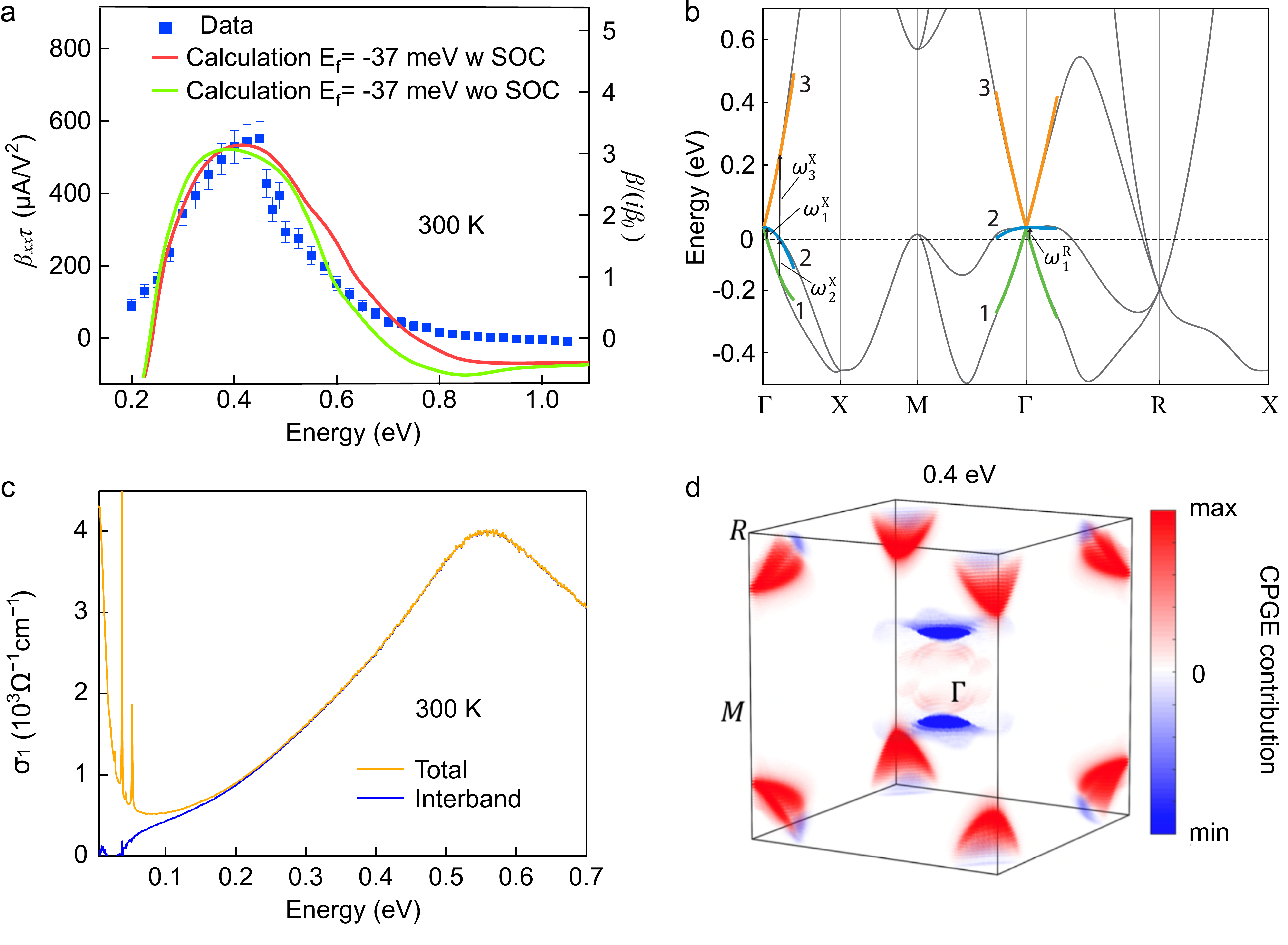}
\caption{\textbf{Room tempreture CPGE spectrum, optical conductivity and band structure for CoSi.} 
\textbf{a},  Measured second-order CPGE photo-conductivity ($\beta_{xx}\tau$) as a function of incident photon energy, and \textit{ab-intio} calculations of the CPGE current with and without spin-orbit coupling at room temperature. 
\textbf{b}, The band structure of CoSi without spin-orbit coupling. We define zero energy at the threefold node at the $\Gamma$ point. The double Weyl node at the $R$ point is at -185 meV. The dashed horizontal line indicates the chemical potential $E_f=-37$ meV in our sample, moderately lower from that obtained by DFT ($E_f=-20$ meV). The band structure of the $k \cdot p$ model is shown in green (band 1), blue (band 2) and orange (band 3) obtained by fitting the \textit{ab-intio} band structure (black) up to quadratic corrections. For the $\Gamma-X$ direction we define $\omega^X_1$, $\omega^X_2$ and $\omega^X_3$ as the minimum energy that allows transitions from band 1 to band 2, the maximum energy that allows transition from band 1 to band 2 and the minimum energy that allows transitions from band 2 to band 3 respectively. We define in the same way $\omega^R_1$ in the $R$ direction ($\omega^R_{2,3}$ fall outside the applicability of the quadratic model).   
\textbf{c}, Total (gold) and interband (blue) optical conductivity of CoSi at 300 K. Total conductivity was adapted from Ref. \onlinecite{xuarXiv2020}.  \textbf{d}, Momentum resolved contributions to the CPGE peak at 0.4 eV in the red curve in \textbf{a}.
} 
\label{Fig3}
\end{figure*}

\textbf{CPGE spectrum in CoSi.} After confirming the longitudinal direction of the CPGE, we now study the amplitude of the CPGE current inside the sample at different incident photon energies.
We use circularly-polarized laser pulses with a duration  50-100 fs and a tunable incident photon energy  from 0.2 eV to 1.1 eV  to generate a CPGE inside of the sample. For the first time, we detected THz emission with incident photon energy below 0.5 eV, comparing with previous measurements~\cite{braunNatComm2016,sotomePNAS2019, reesarxiv2019, siricaPRL2019, gaoNatComm2020}. In order to convert the detected THz electric field into the CPGE current inside the sample, we measured a benchmarking ZnTe sample at the same position at each wavelength immediately after measuring CoSi.  ZnTe is useful as a benchmark due to its relatively flat frequency dependence on the electric-optical sampling coefficient for photon energy below the gap~\cite{Cabrera1985}. {See SI section \textbf{II} for the raw data. We first convert the collected THz electric fields on CoSi and ZnTe from the time domain to the frequency domain by a Fourier transformation. By taking the ratio of the two Fourier transforms of the electric fields and considering the Fresnel coefficient, refractive indices and penetration depth, we obtain the ratio between the CPGE response of CoSi and the optical rectification of ZnTe.  The use of ZnTe circumvents assumptions regarding the incident pulse length, the wavelength dependent focus spot size on the sample, and the calculation of collection efficiency of the off-axis parabolic mirrors (see SI section \textbf{I} C for details). }

\begin{figure*}
\centering
\includegraphics[width=\textwidth]{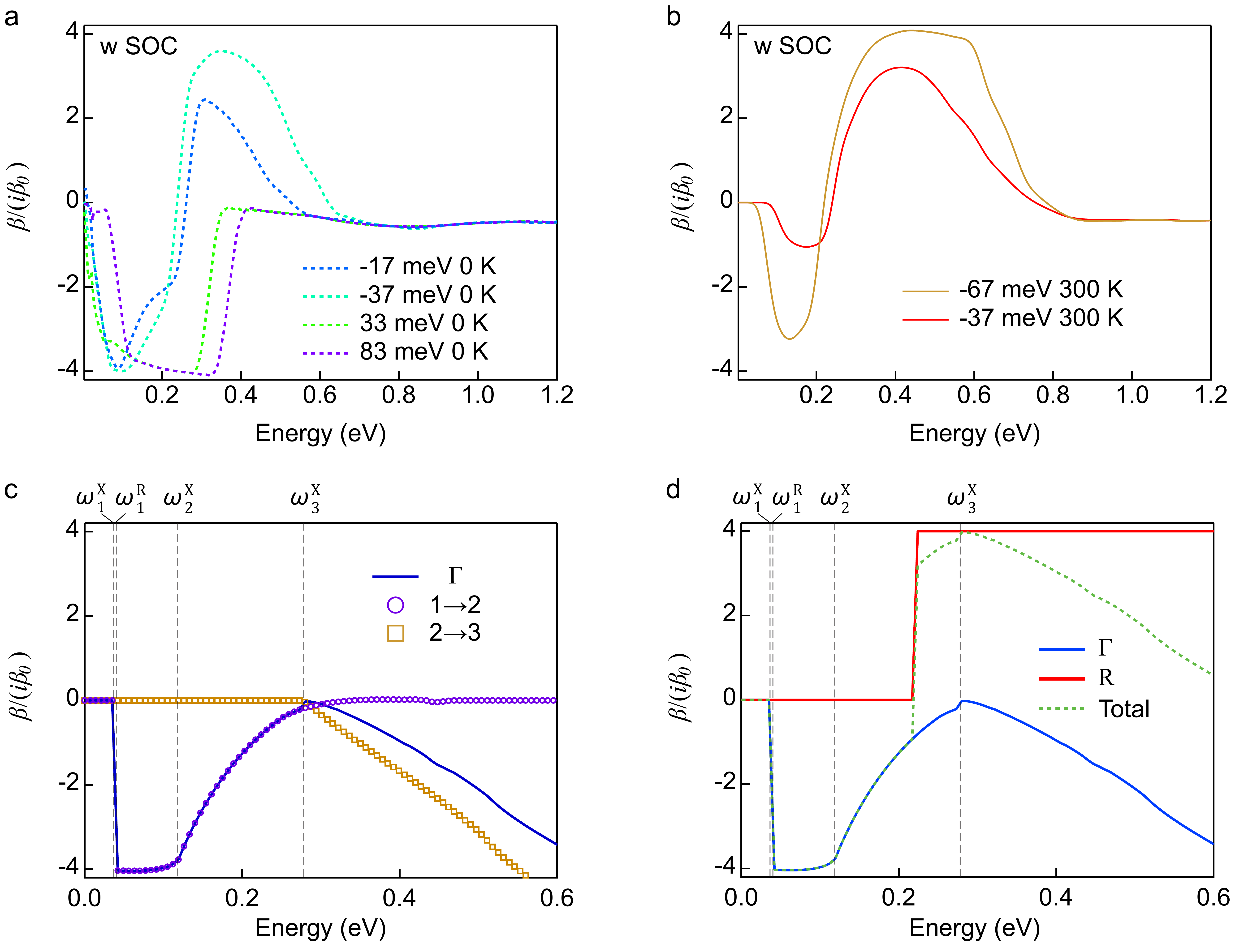}
\caption{\textbf{CPGE calculations for CoSi.} 
\textbf{a,b,} CPGE current obtained by \textit{ab-intio} calculations corresponding to the CoSi band structure with spin-orbit coupling at different chemical potentials at \textbf{a},0 K with 5 meV broadening and \textbf{b}, 300 K with 38 meV broadening. \textbf{c,d,} CPGE current calculation from the $k\cdot p$ model, with parameters $(v,a,b,c) = (1.79,1.07,-1.72,3.26)$. \textbf{c,} The contributions to the CPGE current from transitions near the $\Gamma$ point are shown by open purple circles for transitions from band 1 to 2 and as open gold squares for transitions form band 2 to 3. A solid blue line shows the sum of the contributions form transitions from 1 to 2 and from 2 to 3, which is the total CPGE contribution from the $\Gamma$ point. \textbf{d,} The contribution from the $R$ point to the CPGE is shown by a step of 4 in red. The total contribution to the CPGE from the $\Gamma$ and the $R$ point is shown in green.  A quantized dip/plateau is observed when the frequency is between $\omega^R_1$ and $\omega^X_2$, which allow transitions from band 1 to band 2 only. The dip is determined by transitions around $\Gamma$ only, while the peak has contributions from excitations near the $\Gamma$ and $R$ points.    
} 
\label{Fig4}
\end{figure*}

{The CPGE follows $j(\Omega)=\frac{\beta_{xx}}{i\Omega+1/\tau}E_0^2(\Omega)$, where $\Omega$ is the THz frequency  and $\tau$ is the hot-carrier lifetime. When the hot-carrier lifetime satisfies $\tau\ll1/\Omega$, $j\approx \beta_{xx} \tau E_0^2$. This is the case for the current experiment as $j(\Omega)$ depends weakly on $\Omega$. (See Fig.S6a in SI section \textbf{III}.) {The second-order photo-conductivity plotted in Fig. \ref{Fig3}\textbf{a} was an average value of the CPGE over the frequency range of 0.5 - 2.0 THz in Fig.S6a in SI, which should also be the DC limit.}   When $\tau$ is much longer than the pulse width, which is in the quantization regime, $dj/dt=\beta_{xx}E_0^2$. When $1/\tau$ is comparable to $\Omega$, the CPGE conductivity, $\frac{\beta_{xx}}{i\Omega+1/\tau}$, will have strong frequency dependence in the THz regime, which will enable the extraction of $\beta_{xx}$ and $\tau$ separately.} 
Anticipating our theory analysis, we note that the CPGE spectrum is determined by the only symmetry-independent non-zero CPGE response tensor $\beta_{xx}$, which is a photocurrent rate multiplied by the hot-carrier lifetime  $\tau$. The measured CPGE photocurrent per incident field squared as a function of frequency, which we will denote as the CPGE spectrum, is shown in Fig.~\ref{Fig3}\textbf{a} for room temperature, showing a peak value of 550 ($\pm 45$) $\mu A/V^{2}$  at 0.4 eV.  The CPGE spectrum peak value is much larger than the photo-galvanic effect in any chiral crystals reported in the literature \cite{ZhangPRB2019}, BaTiO$_3$~\cite{feiPRB2020}, single-layer monochalcogenides \cite{rangelPRL2017,feiPRB2020}, the colossal bulk-photovoltaic response in TaAs~\cite{osterhoudtNatMat2019} and RhSi in the same space group~\cite{reesarxiv2019, NiarXiv2020}.\\

\textbf{First-principle calculation.} Next we address the relationship between the large photo-conductivity peak  and the multifold fermions near the Fermi level shown in Fig.~\ref{Fig3}\textbf{b}.  In Fig.~\ref{Fig3}\textbf{a}, we show our ab-initio calculations of the CPGE for CoSi with and without spin-orbit coupling (SOC) at room temperature and at a chemical potential $E_f$ crossing the flat hole band, as indicated by the dashed line in  Fig.~\ref{Fig3}\textbf{b} (see Methods and SI section \textbf{IV}). They quantitatively reproduce the experimental data across a wide frequency range. The SOC splitting, $\approx$ 20 meV at the $\Gamma$ point node, determines the finer structure in the optical response~\cite{xuarXiv2020}.

To match the first-principles calculations with the CPGE spectrum, we considered $\beta_{xx}$, which is related to the CPGE trace by $\beta=3\beta_{xx}$ due to cubic symmetry\cite{dejuanNatComm2017, changPRL2017}. {Note that $\beta_{xx}$ is directly calculated from the band structure at certain chemical potential.} It is plotted in Fig.~\ref{Fig3}\textbf{a} times $\tau$, the only free parameter, which was determined by matching the calculated peak width and magnitude to the CPGE data. This self-consistent constraint is satisfied with a broadening $\hbar/\tau \approx 38$ meV  at 300 K. As shown from the right $y$ axis of Fig.~\ref{Fig3}(a), the CPGE trace $\beta$ reaches 3.3 ($\pm$0.3) in units of the quantization constant $\beta_0$. 

For frequencies below 0.6 eV, all the interband excitations on CoSi occur in the vicinity of its multifold bands at the nodes $\Gamma$ and $R$~\cite{xuarXiv2020}. We conclude this from the optical conductivity on CoSi at 300 K that is shown in Fig.~\ref{Fig3}\textbf{c}. With the subtraction of the Drude response and the four phonon peaks, the interband contribution has a kink at $\sim$ 0.2 eV that separates two quasi-linear regimes. The details of the theoretical and experimental studies could be found in Ref.~\onlinecite{xuarXiv2020}. Below 0.2 eV, the interband excitation involve the threefold fermion at the $\Gamma$ point, while the excitations near the double Weyl fermion at $R$ become active only above 0.2 eV~\cite{xuarXiv2020}. The main contribution to the peak around 0.6 eV comes from the saddle point $M$~\cite{xuarXiv2020}. In addition to the optical conductivity, our momentum resolved calculation (Fig.~\ref{Fig3}\textbf{d}) for the CPGE peak at 0.4 eV reveals that it originates from these multifold fermions, which contribute with opposite signs to the CPGE current. Therefore, this giant CPGE peak has a topological origin, although it is not quantized due to the sum of contributions of two kinds of multifold fermions and the  quadratic contributions to the energy bands.

The position of the chemical potential is crucial to relate the CPGE to quantization. 
Our ab-initio calculations, supported by our low-energy analysis below, reveal that the CPGE shows a dip-peak structure as the one of Fig.~\ref{Fig4}\textbf{a,b}) when $E_{f}$ is below the $\Gamma$ node ($E_{f} < 0$) in our sample. Such Fermi energy is consistent with recent quasi-particle interference~\cite{Yuan19QPI} and linear optical conductivity experiments~\cite{xuarXiv2020}.
Note that this dip-peak structure was clearly observed recently in RhSi
as the energy splitting between the nodes at the $\Gamma$ and $R$ in RhSi is around twice larger than CoSi so that the sign change in CPGE is pushed to around 0.4 eV in RhSi~\cite{NiarXiv2020}. The dip-peak structure for $E_f<0$ is also produced by using a four-band tight binding model for CoSi~\cite{flickerPRB2018} (see Section \textbf{V}). As shown in Fig.\ref{Fig4}\textbf{a}, the dip reaches the quantized value of $4\beta_0$ at low temperatures in the clean limit, and remains quantized for hot carrier lifetime broadening up to 5 meV at 100 meV
photon energy (see SI section \textbf{IV} and Fig. S3 for details.). The quantization of the dip is determined by the threefold fermion at the 
$\Gamma$ point as the vertical excitations at the $R$ point are Pauli blocked below 0.2 
eV~\cite{xuarXiv2020}. {Therefore, the CPGE will not be quantized in the current sample at low temperature in the photon energy range of 0.2 eV - 1.1 eV as the peak around 0.4 eV after the dip appears non-universal in general due to contributions from both nodes at $\Gamma$ and $R$ (see see $E_f=-37$ meV curve in Fig.\ref{Fig4}\textbf{a}).}    However, if $E_f$ is decreased further to lie close to the $R$ point, this peak can reach 4$\beta_0$ at room temperature even with a broadening of $38$ meV (see $E_f=-67$ meV curve in Fig.\ref{Fig4}\textbf{b}). As discussed below, this peak originates from the double Weyl fermion at the $R$ point, and it is enabled by an accidental window of vanishing CPGE contribution from the $\Gamma$ point. Finally, we note that electron-electron interactions can also correct the quantized value, as occurs for chiral Weyl semimetals \cite{Avdoshkin20}. While it is currently unknown 
how relevant these corrections are for multifold fermions,
the large hole and electron pockets at $\Gamma$ and $R$ in CoSi suggest that screening should be strong and therefore interactions should have a small effect. The good agreement of our model calculations with the data, shown in Fig.\ref{Fig3}\textbf{a}, is also consistent with this point of view. Experimentally, from the optical conductivity measurements we estimate a relative dielectric constant $\epsilon_1$ of the order of -2500 at 300 K and -10000 at 10 K, further supporting a normal metallic behavior with very large screening of interactions. Also, specific heat measurements on CoSi also showed that it is a weakly correlated semimetal, as evidenced by a normal metallic Sommerfeld constant \cite{petrovaPRB2010}. Because of these reasons interactions are neglected in this work.\\

\textbf{$k \cdot p$ model.} To understand the origin of the dip-peak structure, it is necessary to describe the curvature of the middle band. To this end we derived a low-energy $k \cdot p$ type model keeping symmetry-allowed terms up to quadratic order in momentum $\vec{k}$. The resulting Hamiltonian reads

\begin{align}
H = v\vec{k}\cdot \vec{S} +\left(\begin{array}{ccc}
c_1 k^2 - 2ck_z^2&  b k_y k_z&  b k_z k_x\\
b k_y k_z& c_1k^2 - 2ck_x^2 & b k_x k_y\\
b k_z k_x& b k_x k_y& c_1k^2 - 2ck_y^2
\end{array}\right),
\end{align}
where $\vec{S}$ is the vector of spin-1 matrices, and $k=|\vec{k}|$.
We fixed its coefficients $v, b$, $c$, and $c_1 = \frac{1}{3}(3a+2c)$ with a fit to the band structure shown in Fig.~\ref{Fig3}\textbf{b} around the $\Gamma$ point. The second term includes three out of the four symmetry-allowed quadratic terms because the fourth has a negligible effect on the CPGE (see SI section \textbf{VI} for details). The energies expanded to second order in momentum for the three bands are plotted as colored lines in Fig.~\ref{Fig3}\textbf{b}. The coefficients $b$ and $c$ determine the curvature in the $\Gamma-X$ and $\Gamma-R$ directions, respectively. For the $R$ point bands, we use a spin-degenerate double Weyl model that has a step increase in the CPGE current by $4\beta_0$ when excitations at $R$ are allowed in Fig.~\ref{Fig4}\textbf{d}~\cite{changPRL2017}.

The possible optical transitions in the band structure near the $\Gamma$ point are illustrated in  Fig.~\ref{Fig3}\textbf{b}. We label the bands with increasing energies as 1,2,3. For $E_f$ above the threefold node, the only possible transition is from bands 2 to 3. As the frequency increases, this transition becomes active and yields a monotonically increasing joint density of states (JDOS)\cite{flickerPRB2018}. As shown in Fig.~\ref{Fig4}
\textbf{c}, for $E_f$ below the node, however, two types of transitions contribute: 1 to 2 and 2 to 3. The first transition from band 1 to band 2 (open purple circles) is active for a small range of energies, and then decays to zero. The second transition from band 2 to band 3 (open gold squares) only starts picking up at larger frequencies, leaving a dip in the JDOS and, therefore, a dip in the CPGE (solid blue line). The different frequencies where the transitions become active or inactive are labeled in Fig.\ref{Fig3}\textbf{b} and Fig.\ref{Fig4}\textbf{c,d}. Fig.\ref{Fig4}\textbf{d} show that when we add the contributions from the threefold fermions at $\Gamma$ and double Weyl fermion at $R$, the existence of the dip from the threefold fermions leads to the dip-peak structure observed in the ab-initio calculations, only when $E_f$ is  below the threefold node. In the $k\cdot p$ model, we also show that while the dip is universally quantized, the peak is not because of the incomplete transitions from $\Gamma$. {Note that the quantization of the peak not only depends on the $\Gamma$ contribution but it also may be altered by the quadratic dispersion of the double Weyl fermion when it fully contributes to CPGE.} However, as shown in Fig. \ref{Fig4}\textbf{b}, decreasing $E_f$ further could be used to diminish the contribution from the threefold fermions at around 0.4 eV and reveal the quantization due to the $R$ point (see SI section \textbf{VI} for details).\\

\textbf{Discussion} By studying the CPGE in the chiral topological semimetal CoSi we found a large longitudinal photo-conductivity in the mid-infrared regime, which has a topological origin linked to the existence of multifold fermions in this material. CoSi could potentially be used as a new mid-infrared detector based on a topological mechanism if the hot-carrier lifetime could be increased to around 1 ps, as observed in other semimetals \cite{zhuAPL2017}.
Moreover, our theory suggests that a quantized CPGE is within reach in CoSi by several means. With the chemical potential below the threefold node, the very narrow quantized plateau around 100 meV corresponding to the $\Gamma$ node could be accessible at low temperatures if the hot-carrier lifetime increases by one order of magnitude. Also, electron doping the $E_f$ above the threefold node will result in a wider quantized plateau over 100-350 meV from the $\Gamma$ node corresponding to the dip we calculate at low temperatures also if the hot-carrier lifetime increases by one order of magnitude. A quantized plateau around 0.4 eV corresponding to the $R$ node can be revealed at $T$=300 K by hole-doping, even with a similar short hot carrier lifetime as that of our current sample. We expect that these possibilities, opened by our work, motivate further effort on crystal growth with longer hot carrier lifetime and different doping, as well as infrared time-resolved measurements 
in the mid-infrared regime to probe the hot carrier dynamics. The methods developed in our work  could also be applied to other chiral topological semimetals \cite{bradlynScience2016,changNatMat2018} to realize the quantized CPGE.

\section{Methods}

\textbf{Crystal growth}
High quality CoSi single crystals were prepared by a high temperature flux method with tellurium as flux.  Cobalt pieces (Alfa Aesar 99.98$\%$), silicon pieces (Alfa Aesar 99.999$\%$), and tellurium lumps (Alfa Aesar 99.99$\%$) with the molar ratio of 1:1:15-20 were set in an alumina crucible and then sealed in a fused silica ampule in around 0.8 atm argon environment.  The ampule was heated to 1100 $^{\circ}$C with a speed of about 150 $^{\circ}$C/hour.  After soaking at 1100 $^{\circ}$C for 10 hours, the ampule was cooled down to 700 $^{\circ}$C at the rate of 2 $^{\circ}$C/hour, and the excess flux was centrifuged out at that temperature to get several single crystals with large (111) facet\cite{xuXPRB2019}. Crystals larger than 2 mm $\times$ 2 mm were picked for THz emission experiments. We also performed a spatial SHG scanning of the sample and found an homogeneous signal. 

\textbf{Second harmonic generation fit}

\noindent For CoSi (111), the fits are:

\begin{equation}
I_{parallel}(\theta)=\frac{1}{6}|\chi^{(2)}_{xyz}|^2(cos^3\theta-3\cos\theta\sin^2\theta)^2.
\end{equation}

\begin{equation}
I_{crossed}(\theta)=\frac{1}{6}|\chi^{(2)}_{xyz}|^2(\sin^3\theta-3\cos^2\theta\sin\theta)^2.
\end{equation}

$\chi^{(2)}_{xyz}$ is the only non-zero SHG tensor element in CoSi. $\theta$ is the angle between the incident polarization and the [1,1,-2] axis.

\textbf{Terehertz emission spectroscopy}
A laser beam from a Ti:sapphire  amplifier (center photon energy 1.55 eV, repetition rate 1 kHz, duration $\sim$ 35 fs) was split by a beam splitter into pump and probe beams. On the pump side, an optical parametric amplifier  is used to convert the photon energy to 0.47-1.1 eV (pulse duration 40-70 fs), and a different frequency generation  is used to further convert photon energy to 0.20-0.48 eV (pulse duration 70-110 fs). The laser beams were then focused by a 40-cm BaF$_2$ lens or a 40-cm germanium lens onto the sample with a diameter of ~1 mm under 45 degree angle of incidence. A typical pump power of 15 $\mu$J per pulse was used, which falls into the linear response range. The emitted THz  wave was collected by an off-axis parabolic mirror (OAP) and focused by another OAP onto an electro-optic (EO) crystal, ZnTe (110). The probe beam was co-propagating with the THz wave into the EO crystal to detect the THz electric field using EO-sampling method\cite{shan2004terahertz}. All of the measurement were performed in a dry-air environment with relative humidity less than 3$\%$ to avoid water absorption. To control the polarization of pump pulses, a quartz-MgF$_2$ achromatic quarter-wave plate (600-2700 nm, retardance error $\leq$ $\lambda$/500) and a MgF$_2$ achromatic quarter-wave plate (2500-7000 nm, retardance error $\leq$ $\lambda$/100) were used. A THz wire-grid polarizer was used to extract out-of-plane (E$_y$) and in-plane (E$_{xz}$) components of THz electric field.
A benchmarking crystal ZnTe (110) was used as a standard candle to extract the photogalvanic response from CoSi. Both crystals were mounted on a computer-controlled motor to reliably change the position. For each incident photon energy, measurement of CoSi was immediately followed by ZnTe to avoid long-term fluctuation of laser power. By comparing the THz electric field of CoSi and ZnTe in frequency domain, the photogalvanic response of the CoSi crystal could be precisely obtained (see SI section \textbf{III} for details).

\textbf{First principle calculation}
To calculate the CPGE current, we obtain the density-functional theory (DFT) Bloch wave functions from the Full-Potential Local-Orbital program (FPLO) \cite{koepernik1999full} within the generalized gradient approximation (GGA) \cite{perdew1996}. By projecting the Bloch wave functions onto Wannier functions, we obtain a tight-binding Hamiltonian with 104 bands from 3$d$, 4$s$, 4$p$ orbitals of Co and 3$s$, 3$p$ orbitals of Si, which we use for efficient evaluation of the CPGE photocurrent.

To implement the CPGE integrals in Eq. \eqref{eq:injection}, the Brillouin zone was sampled by $k$-grids from $200\times200\times200$ to $960\times960\times960$ \cite{zhang2018photogalvanic}. Satisfactory convergence (less than 2\% change) was achieved for a k-grid of size $400\times400\times400$. The temperature dependence is implemented by the Fermi-Dirac distribution function and we also include a hot-carrier lifetime broadening factor (see SI section \textbf{IV} for details). CoSi is in space group $P2_13$ ($\#$198), with point group $23$ ($T$). Owing to the two-fold glide rotation symmetry $s_{2x}, s_{2y}, s_{2z}$, only diagonal tensor elements are nonzero, and the $C_3$ rotation symmetry further leads to a single independent component $\beta_{xx}=\beta_{yy}=\beta_{zz}$. We carefully checked the symmetry of numerically calculated tensor elements with the tensor shape given by lattice symmetry and found the errors to be  within $10^{-6}$. The full circular photogalvanic effect tensor is given by

\begin{eqnarray}
	\nonumber
{\beta_{ab}}(\omega) &=& {\frac{i\pi e^3}{4\hbar}}  \int_{\rm BZ} \frac{dk}{(2\pi)^2} {\sum_{n>m}} \epsilon^{bcd} f_{nm}\\
&\times& \Delta^a_{mn}{\rm Im} [r^d_{nm}r^c_{mn}] \mathcal{L}_\tau(E_{nm}-\hbar\omega), \label{eq:injection}
\end{eqnarray}
where $E_{mn} \equiv E_m -E_n$, $f_{mn} \equiv f_m - f_n$ are the difference of band dispersion and Fermi-Dirac distribution respectively, $\Delta^a_{mn}\equiv\partial_{k_a}E_{mn}/\hbar$, $r_{mn}^a \equiv i \langle m|\partial_{k_a}n\rangle$ is the interband transition matrix element or off-diagonal Berry connection. The finite relaxation time $\tau$ is considered via the Lorentzian function $\mathcal{L}_\tau(E_{nm}-\omega)$.

\section{addendum}
We thank C. L. Kane for helpful discussions and N. P. Armitage and J. Stensberg for proof-reading the manuscript. Z.N. and L.W. are supported by ARO YIP award under the Grant W911NF1910342. X.H. is  partially supported by ARO MURI award under the Grant W911NF2020166. The acquisition of the oscillator laser for the SHG experiment is supported by NSF through the University of Pennsylvania Materials Research Science and Engineering Center (MRSEC) (DMR-1720530). E.J.M's theoretical work is supported by the DOE under grant DE FG02 84ER45118. Research at the University of Maryland was supported by the Gordon and Betty Moore Foundation's EPiQS Initiative through Grant No. GBMF9071, and the Maryland Quantum Materials Center. Y.Z. is currently supported by the the DOE Office of Basic Energy Sciences under Award desc0018945 to Liang Fu.     F. J. acknowledges funding from the Spanish MCI/AEI through grant No. PGC2018-101988-B-C21. O.P. is supported by an FPU predoctoral contract from MECD No. FPU16/05460 and the Spanish grant PGC2018- 099199-BI00 from MCIU/AEI/FEDER. B.X. is supported by the Schweizerische Nationalfonds (SNF) by Grant No. 200020-172611. A. G. G. is supported by the ANR under the grant ANR-18-CE30-0001-01 (TOPODRIVE) and the European Union Horizon 2020 research and innovation programme under grant agreement No. 829044 (SCHINES).  Y. Z, K. M. and C. F. acknowledge the financial support from the European Research Council (ERC) Advanced Grant No. 742068 ``TOP-MAT''; Deutsche Forschungsgemeinschaft (DFG) through SFB 1143, and the Würzburg-Dresden Cluster of Excellence on Complexity and Topology in Quantum Matter-ct.qmat (EXC 2147, Project No. 390858490). The DFT calculations are carried on Draco cluster of MPCDF, Max Planck society.

\textit{Competing Interests: }The authors declare that they have no
competing financial interests.

\textit{Data availability:} All data needed to evaluate the conclusions in the paper are present in the paper and/or the Supplementary Information. Additional data related to this paper could be requested from the authors.

\textit{Correspondence: }Correspondence and requests for materials
should be addressed to L.W. (liangwu@sas.upenn.edu) \\

\section{Author Contribution}
L.W. conceived the project and coordinated the experiments and theory. Z.N., X.H. and L.W. built the THz emission setup. Z.N. performed the THz emission experiments.  Z.N., and L.W. analyzed the data.  Z.N., L.W. and E.M. performed the symmetry analysis. B.X. performed the optical conductivity measurement. Y.Z. performed DFT calculation. O.P. and F.J. performed the $k \cdot p$ calculation. A.G.G. performed the tight-binding calculation. K.W., K.M., J.P., and C.F. grew the crystals. L.W., F.J.,and A.G.G.  wrote the manuscript from contributions of all authors. L.W., Z.N., O.P., A.G.G., F.J. and Y. Z. wrote the supplementary information. E.M. edited the manuscript.
Z.N., K.W. and Y.Z. contributed equally to this work.

\begin{widetext}

\newpage
\newpage

\onecolumngrid
\begin{center}
  \textbf{\large Supplementary Information for ``Giant topological longitudinal circular photo-galvanic effect in the chiral multifold semimetal CoSi"}\\[.2cm]\

\end{center}

\setcounter{equation}{0}
\setcounter{figure}{0}
\setcounter{table}{0}
\setcounter{page}{1}
\setcounter{section}{0}
\renewcommand{\theequation}{S\arabic{equation}}
\renewcommand{\thefigure}{S\arabic{figure}}
\renewcommand{\bibnumfmt}[1]{[S#1]}
\renewcommand{\citenumfont}[1]{S#1}

\section{Conversion of THz signal}

As will be explained in the following sections, the symmetry of CoSi only allows CPGE current along the wave vector direction inside the material determined by the incident wave vector. Normal incident laser pulses can only generate longitudinal ultrafast current, which is perpendicular to the surface and does not emit THz radiation into the free space. Therefore, in order to make THz pulses radiate out to the free space in the far-field, we choose the incident angle to be at 45 degrees in the reflection geometry. The schematic diagrams of the air-sample interface of both the benchmarking crystal ZnTe and the sample CoSi for the THz emission process are shown in Fig. \ref{geometry}. The incident light (red) gets refracted when entering the sample at the angle of refraction $\theta_{in}$. The ultrafast excitation light inside the sample generates an ultrafast polarization (yellow) in ZnTe or an ultrafast current (yellow) in CoSi, both of which emit THz radiation. Note the current drawn in CoSi contains both CPGE and LPGE, so its direction is not strictly along the light vector direction. Only if CPGE is considered, the current direction is along the light vector direction due to the cubic symmetry of CoSi. In the process of THz emission, only the component of polarization or current that is perpendicular to the in-phase THz radiation direction (blue) generates the coherent THz radiation in the far-field. The direction of the in-phase THz radiation ($\theta_{out}$ off the interface normal) is determined by the refractive index at the THz range.

\subsection{THz generation from ZnTe}
A ZnTe crystal generates THz emission via optical rectification. The nonlinear polarization $P$ under the driving electric field $E$ is characterized by second-order susceptibility tensor $\chi^{(2)} (0;\omega,\omega)$,

\begin{equation}
\mathbf{P}(0)=\epsilon_0\chi^{(2)}(0;\omega,\omega):\mathbf{E\ E^*}.
\end{equation}

ZnTe belongs to space group $F\overline{4}3m$, of which $\chi^{(2)}$ contains only one independent parameter,
\begin{equation}
\chi^{(2)}=\left(
\begin{array}{ccc}
\left(
\begin{array}{c}
0 \\
0 \\
0 \\
\end{array}
\right) & \left(
\begin{array}{c}
0 \\
0 \\
\chi_{41} \\
\end{array}
\right) & \left(
\begin{array}{c}
0 \\
\chi_{41} \\
0 \\
\end{array}
\right) \\
\left(
\begin{array}{c}
0 \\
0 \\
\chi_{41} \\
\end{array}
\right) & \left(
\begin{array}{c}
0 \\
0 \\
0 \\
\end{array}
\right) & \left(
\begin{array}{c}
\chi_{41} \\
0 \\
0 \\
\end{array}
\right) \\
\left(
\begin{array}{c}
0 \\
\chi_{41} \\
0 \\
\end{array}
\right) & \left(
\begin{array}{c}
\chi_{41} \\
0 \\
0 \\
\end{array}
\right) & \left(
\begin{array}{c}
0 \\
0 \\
0 \\
\end{array}
\right) \\
\end{array}
\right)
\end{equation}

We use [1 1 0] cut ZnTe as a THz emission reference. To simplify the following calculation, we rotate $\chi^{(2)}$ to align the [1 1 0] direction of crystal along the $z$ axis in the lab frame,

\begin{figure}

\centering
\includegraphics[width=0.75\textwidth]{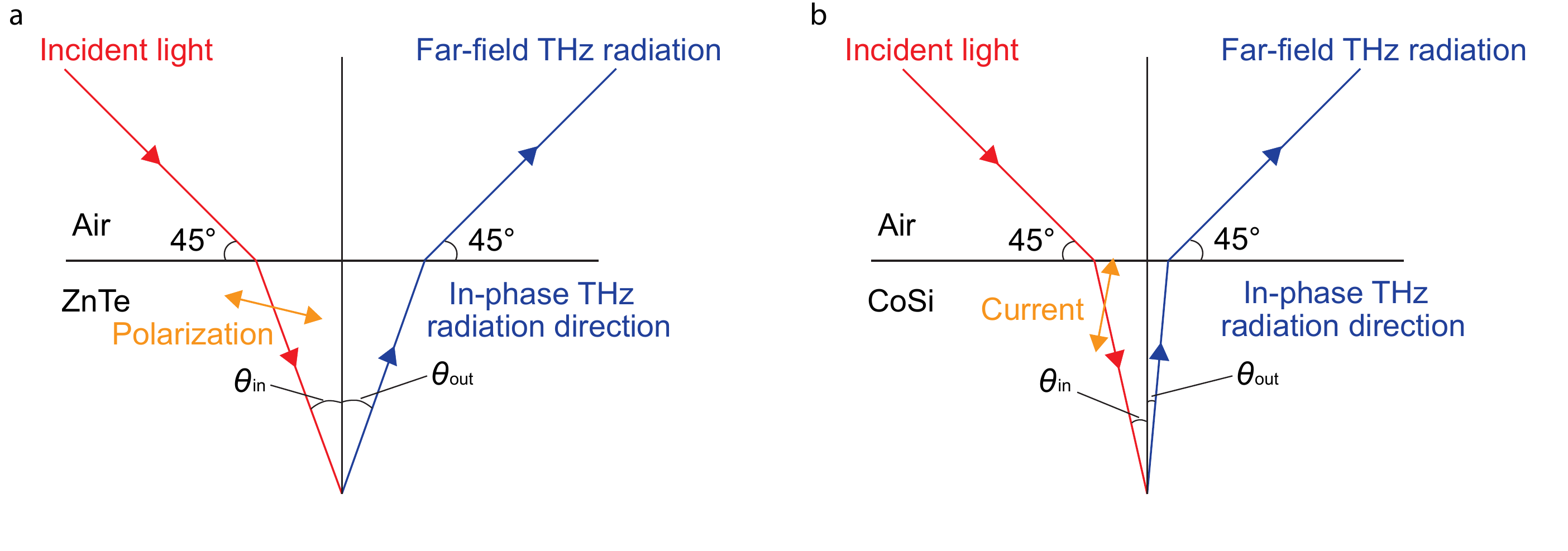}
\caption{{a, Geometry of the air/sample interface of ZnTe. b, Geometry of the air/sample interface of CoSi. Red arrows represent the direction of excitation light, and the blue arrows represent the direction of in-phase THz wave. The yellow arrows represent the polarization from the optical rectification effect in ZnTe and the current from the photogalvanic effect in CoSi. Note the current in b contains both CPGE and LPGE contribution. The CPGE contribution is strictly along the direction of the excitation light inside the sample, while the LPGE contribution is not restricted.}}
\label{geometry}
\end{figure}
\begin{equation}
\chi^{(2)}=\left(
\begin{array}{ccc}
\left(
\begin{array}{c}
\frac{3 \chi_{41} (\sin (2 \phi )+1) (\cos (\phi )-\sin (\phi ))}{2 \sqrt{2}} \\
\frac{\chi_{41} (3 \sin (2 \phi )-1) (\sin (\phi )+\cos (\phi ))}{2 \sqrt{2}} \\
0 \\
\end{array}
\right) & \left(
\begin{array}{c}
\frac{\chi_{41} (3 \sin (2 \phi )-1) (\sin (\phi )+\cos (\phi ))}{2 \sqrt{2}} \\
-\frac{\chi_{41} (3 \sin (2 \phi )+1) (\cos (\phi )-\sin (\phi ))}{2 \sqrt{2}} \\
0 \\
\end{array}
\right) & \left(
\begin{array}{c}
0 \\
0 \\
\frac{\chi_{41} (\sin (\phi )-\cos (\phi ))}{\sqrt{2}} \\
\end{array}
\right) \\
\left(
\begin{array}{c}
\frac{\chi_{41} (3 \sin (2 \phi )-1) (\sin (\phi )+\cos (\phi ))}{2 \sqrt{2}} \\
-\frac{\chi_{41} (3 \sin (2 \phi )+1) (\cos (\phi )-\sin (\phi ))}{2 \sqrt{2}} \\
0 \\
\end{array}
\right) & \left(
\begin{array}{c}
-\frac{\chi_{41} (3 \sin (2 \phi )+1) (\cos (\phi )-\sin (\phi ))}{2 \sqrt{2}} \\
\frac{3 \chi_{41} (\sin (\phi )+\cos (\phi )) (1-2 \sin (\phi ) \cos (\phi ))}{2 \sqrt{2}} \\
0 \\
\end{array}
\right) & \left(
\begin{array}{c}
0 \\
0 \\
-\frac{\chi_{41} (\sin (\phi )+\cos (\phi ))}{\sqrt{2}} \\
\end{array}
\right) \\
\left(
\begin{array}{c}
0 \\
0 \\
\frac{\chi_{41} (\sin (\phi )-\cos (\phi ))}{\sqrt{2}} \\
\end{array}
\right) & \left(
\begin{array}{c}
0 \\
0 \\
-\frac{\chi_{41} (\sin (\phi )+\cos (\phi ))}{\sqrt{2}} \\
\end{array}
\right) & \left(
\begin{array}{c}
\frac{\chi_{41} (\sin (\phi )-\cos (\phi ))}{\sqrt{2}} \\
-\frac{\chi_{41} (\sin (\phi )+\cos (\phi ))}{\sqrt{2}} \\
0 \\
\end{array}
\right) \\
\end{array}
\right),
\end{equation}
where we have a free parameter $\phi$ since the crystal can rotate freely along the $z$-axis and $\phi$ is the angle between crystal direction [1 -1 0] and the $x$-axis in the lab.

For all the experiments in this paper, we make $\phi=0$, align the polarization of incident light along the $x$-axis, and detect THz wave component in the $xz$-plane, so

\begin{equation}
\chi^{(2)}=\left(
\begin{array}{ccc}
\left(
\begin{array}{c}
\frac{3 \chi_{41}}{2 \sqrt{2}} \\
-\frac{\chi_{41}}{2 \sqrt{2}} \\
0 \\
\end{array}
\right) & \left(
\begin{array}{c}
-\frac{\chi_{41}}{2 \sqrt{2}} \\
-\frac{\chi_{41}}{2 \sqrt{2}} \\
0 \\
\end{array}
\right) & \left(
\begin{array}{c}
0 \\
0 \\
-\frac{\chi_{41}}{\sqrt{2}} \\
\end{array}
\right) \\
\left(
\begin{array}{c}
-\frac{\chi_{41}}{2 \sqrt{2}} \\
-\frac{\chi_{41}}{2 \sqrt{2}} \\
0 \\
\end{array}
\right) & \left(
\begin{array}{c}
-\frac{\chi_{41}}{2 \sqrt{2}} \\
\frac{3 \chi_{41}}{2 \sqrt{2}} \\
0 \\
\end{array}
\right) & \left(
\begin{array}{c}
0 \\
0 \\
-\frac{\chi_{41}}{\sqrt{2}} \\
\end{array}
\right) \\
\left(
\begin{array}{c}
0 \\
0 \\
-\frac{\chi_{41}}{\sqrt{2}} \\
\end{array}
\right) & \left(
\begin{array}{c}
0 \\
0 \\
-\frac{\chi_{41}}{\sqrt{2}} \\
\end{array}
\right) & \left(
\begin{array}{c}
-\frac{\chi_{41}}{\sqrt{2}} \\
-\frac{\chi_{41}}{\sqrt{2}} \\
0 \\
\end{array}
\right) \\
\end{array}
\right),
\end{equation}

\begin{equation}
\mathbf{E}=E_0\begin{pmatrix}
t_p\cos(\theta_{in})\\
0\\
t_p\sin(\theta_{in})
\end{pmatrix},
\end{equation}

\begin{equation}
P_{eff}=\mathbf{P} \cdot\begin{pmatrix}\cos(\theta_{out})\\ 0\\ \sin(\theta_{out})\end{pmatrix},
\end{equation}
where $t_p$ is the Fresnel coefficient, $\theta_{in}$ is the incident angle inside the sample, $\theta_{out}$ is the refracted angle of the THz light inside the sample {(See Fig. \ref{geometry})}, and $P_{eff}$ is the component of $\mathbf{P}$ perpendicular to the detecting direction. Combing (S1)(S4)(S5)(S6), we get
\begin{equation}
P_{eff}= \frac{ 4 \sin (2 \theta_{in}) \sin(\theta_{out})+5 \cos 2 (\theta_{in} )\cos (\theta_{out})+ \cos (\theta_{out})}{4 \sqrt{2}} t_p^2 \epsilon_0 \chi_{41}E_0^2\equiv\gamma^{ZnTe}\epsilon_0 \chi_{41}E_0^2.
\end{equation}

Here, $P_{eff}$ is in the time domain and has a pulsed shape with a time scale of picosecond, so its Fourier transform $P_{eff}(\Omega)$ is in the THz range. The generated THz field strength just outside the sample surface in the frequency domain can be written as
\begin{equation}
E_{eff}^{ZnTe}(\Omega)=\frac{Z_0\Omega P_{eff}(\Omega)}{2 n^{ZnTe}(\Omega)} T_p^{ZnTe} l^{ZnTe}(\Omega),
\end{equation}
where $\Omega$ is THz frequency, $Z_0$ is the impedance of free space, $P_{eff}(\Omega)$ is the Fourier transform of $P_{eff}$, $n^{ZnTe}(\Omega)$ is the refractive index of ZnTe for frequency $\Omega$, $T_p^{ZnTe}$ is the Fresnel coefficient, and $l^{ZnTe}$ is the coherent length of the THz wave in ZnTe.

\subsection{THz generation from CoSi}
CoSi belongs to the $P2_13$ space group, and its second-order conductivity tensor has the same form as (S2) except that the parameter is a complex number $\sigma+i\eta$. In the experiment, samples with (111) surface are measured. After a rotation transformation of the conductivity tensor, the total current $\mathbf{j}$ can be written as
\begin{equation}
\mathbf{j}=\sigma^{(2)} :\mathbf{E\ E^*}
\label{jee}
\end{equation}
with
\begin{equation}
\sigma^{(2)}=\left(
\begin{array}{ccc}
\left(
\begin{array}{c}
\sqrt{\frac{2}{3}} \sigma \cos (3 \phi ) \\
\sqrt{\frac{2}{3}} \sigma \sin (3 \phi ) \\
-\frac{\sigma }{\sqrt{3}} \\
\end{array}
\right) & \left(
\begin{array}{c}
\sqrt{\frac{2}{3}} \sigma \sin (3 \phi ) \\
-\sqrt{\frac{2}{3}} \sigma \cos (3 \phi ) \\
i \eta \\
\end{array}
\right) & \left(
\begin{array}{c}
-\frac{\sigma }{\sqrt{3}} \\
-i \eta \\
0 \\
\end{array}
\right) \\
\left(
\begin{array}{c}
\sqrt{\frac{2}{3}} \sigma \sin (3 \phi ) \\
-\sqrt{\frac{2}{3}} \sigma \cos (3 \phi ) \\
-i \eta \\
\end{array}
\right) & \left(
\begin{array}{c}
-\sqrt{\frac{2}{3}} \sigma \cos (3 \phi ) \\
-\sqrt{\frac{2}{3}} \sigma \sin (3 \phi ) \\
-\frac{\sigma }{\sqrt{3}} \\
\end{array}
\right) & \left(
\begin{array}{c}
i \eta \\
-\frac{\sigma }{\sqrt{3}} \\
0 \\
\end{array}
\right) \\
\left(
\begin{array}{c}
-\frac{\sigma }{\sqrt{3}} \\
i \eta \\
0 \\
\end{array}
\right) & \left(
\begin{array}{c}
-i \eta \\
-\frac{\sigma }{\sqrt{3}} \\
0 \\
\end{array}
\right) & \left(
\begin{array}{c}
0 \\
0 \\
\frac{2 \sigma }{\sqrt{3}} \\
\end{array}
\right) \\
\end{array}
\right),
\end{equation}
where the operators $\sigma$ and $\eta$ describe LPGE and CPGE contributions, respectively, and $\phi$ is the angle between the sample axis [2 -1 -1] and x axis in the lab frame.
For most of our experiments, we use only one QWP to change the polarization of the incident light and fix $\phi$=90 degrees. The electric field after the QWP is
\begin{equation}
\mathbf{E}=E_0\begin{pmatrix}
t_p\cos(\theta_{in})\left(\cos^2(\theta)+i \sin^2(\theta)\right)\\
(1-i)t_s\sin(\theta)\cos(\theta)\\
t_p\sin(\theta_{in})\left(\cos(\theta)+i \sin^2(\theta)\right)
\end{pmatrix}
\end{equation}
With \ref{jee} and the geometry shown in Fig. \ref{geometry}, one can calculate the current component coupled to the free space THz radiation in the far field,

\begin{equation}
\begin{aligned}
j_{eff}^{xz}(\theta)= &
i \eta E_0^2 \sin (2 \theta ) \sin \left(\theta _{\text{in}}+\theta _{\text{out}}\right) \left(\Re({t_s t_p^*})+\Im(t_st_p^*)\cos (2\theta)\right)\\&
- \frac{ \sigma E_0^2}{8\sqrt{3}}(\cos (4 \theta )+3) \left| t_p\right| {}^2 \left(2 \sin \left(2 \theta _{\text{in}}\right) \cos \left(\theta _{\text{out}}\right)+\left(1-3 \cos \left(2 \theta _{\text{in}}\right)\right) \sin \left(\theta _{\text{out}}\right)\right)\\&
- \frac{ \sigma E_0^2}{2\sqrt{3}} \sin (2 \theta ) \left(\sin (2 \theta ) \left| t_s\right| {}^2 \sin \left(\theta _{\text{out}}\right)+\sqrt{2} \cos \left(\theta _{\text{in}}\right) \cos \left(\theta _{\text{out}}\right) \left((\cos (2 \theta )+i) t_p t_s^*+(\cos (2 \theta )-i) t_s t_p^*\right)\right)
\label{jeff}
\end{aligned}
\end{equation}

{The parameters $t_s t_p^*$, $\theta_{in}$ and $\theta_{out}$ used in this equation can be determined by the linear optical conductivity of CoSi, which is shown in the main text Fig. 3c. In Fig. \ref{parameters} we show $t_s t_p^*$ (Fig. \ref{parameters}e) and $\theta_{in}$(Fig. \ref{parameters}f) as a funcition of excitation energy and $\theta_{out}$ (Fig. \ref{parameters}h) as a function of radiation frequency (See details in the Sec. \ref{secparameters}).}

\begin{figure}
\centering
\includegraphics[width=\textwidth]{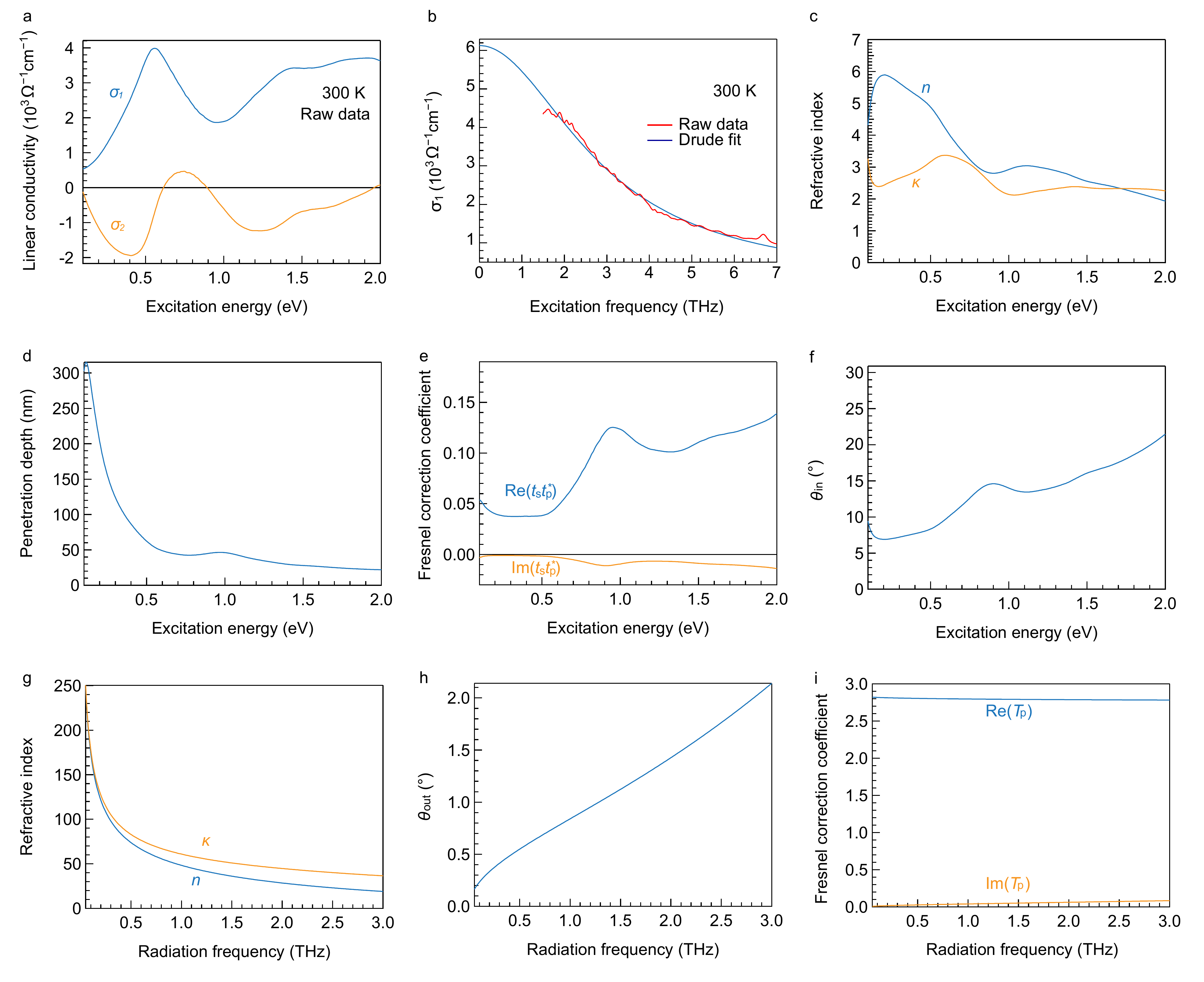}
\caption{{Parameters of CoSi related with the optical  conductivity. a, Optical conductivity of CoSi as a function of excitation energy measured at 300 K. The data is reproduced from Ref. \cite{xuarXiv2020}. Only 0.1-2.0 eV is shown. b, Drude fit (blue) of the optical conductivity raw data (red) at THz range. c-f, Refractive index (c), penetration depth (d), Fresnel correction coefficient (e) and $\theta_{in}$ (f) of CoSi as a function of the excitation energy. g-i, Refractive index (g), $\theta_{out}$ (h) and Fresnel coefficient (i) of CoSi as a function of the THz radiation frequency. These curves are derived based on the optical conductivity following Drude formula fitted in b.}}
\label{parameters}
\end{figure}

Considering $\Re({t_s t_p^*})\gg\Im(t_st_p^*)$ and $\theta_{out}\ll1$ , \ref{jeff} can be further simplified to
\begin{equation}
\begin{aligned}
j_{eff}^{xz}(\theta)=&
\eta E_0^2 \sin (2 \theta ) \sin \left(\theta _{\text{in}}+\theta _{\text{out}}\right) \Re({t_s t_p^*})\\&
- \frac{ \sigma E_0^2}{4\sqrt{3}}(\cos (4 \theta )+3) \left| t_p\right| {}^2 \sin \left(2 \theta _{\text{in}}\right) \cos \left(\theta _{\text{out}}\right)\\&
- \frac{ \sigma E_0^2}{\sqrt{6}} \sin (4 \theta ) \cos \left(\theta _{\text{in}}\right) \Re({t_s t_p^*}).
\end{aligned}
\end{equation}
To get pure CPGE, we measure the current difference between $\theta$=45 degrees and $\theta$=135 degrees,
\begin{equation}
j_{eff}^{CPGE}=\frac{1}{2}\left(j_{eff}^{xz}(\frac{3\pi}{4})-j_{eff}^{xz}(\frac{\pi}{4})\right)= \eta \Re\left(t_s t_p\right) \sin \left(\theta _{\text{in}}+\theta _{\text{out}}\right)E_0^2=\gamma^{CoSi}\eta E_0^2.
\end{equation}
Note the above equation is true for any crystal orientation, i.e., any $\phi$ and surface index.

We then calculate the generated current out of the incident plane, with the same approximation $\Re({t_s t_p^*})\gg\Im(t_st_p^*)$ and $\theta_{out}\ll1$,
\begin{equation}
j_{eff}^{y}(\theta)=-\frac{\sigma E_0^2}{4 \sqrt{3}} \left(\sqrt{2} \left | tp \right |^2 (\cos (4 \theta )+3) \cos ^2(\theta_{in})+2 t_st_p^* \sin (4 \theta ) \sin (\theta_{in})+\sqrt{2} \left |t_s \right |^2 (\cos (4 \theta )-1)\right).
\end{equation}

Note this result does not have any term with $\eta$, meaning CPGE does not produce out-of-plane photocurrent, which is consistent with the longitudinal response. In the main text, we write: $E_{xz}(\theta)=A\sin(2\theta)+B_2\sin(4\theta)+C_2\cos(4\theta)+D_2$ and $E_y(\theta)=B_1\sin(4\theta)+C_1\cos(4\theta)+D_1$. Here $E_{xz}$ is related with $j_{eff}^{xz}$ and $E_{y}$ is related with $j_{eff}^{y}$, and one can see that coefficients $A$,$B_1$,$B_2$,$C_1$,$C_2$,$D_1$,$D_2$ are determined by the CPGE conductivity $\eta$ and LPGE conductivity $\sigma$.

Similar to ZnTe, the generated THz field of CoSi in the near field is
\begin{equation}
E_{eff}^{CoSi}(\Omega)=\frac{Z_0j_{eff}(\Omega)}{2 n^{CoSi}(\Omega)} T_p^{CoSi} l^{CoSi}(\Omega),
\end{equation}
where $j_{eff}(\Omega)$ is the Fourier transform of $j_{eff}$, $n^{CoSi}(\Omega)$ is the refractive index of CoSi, $T_p^{CoSi}$ is the Fresnel coefficient, and $l^{CoSi}$ is the penetration depth of the incident light in CoSi. {Note $n^{CoSi}(\Omega)$, $T_p^{CoSi}$ and $l^{CoSi}$ are determined by the optical conductivity of CoSi and their dependence on radiation frequency or excitation energy are shown in Fig. \ref{parameters} (See details in the Sec. \ref{secparameters}).}

\subsection{Derivation of CoSi response by the benchmarking ZnTe}

The emitted THz wave is collected by two off-axis parabolic mirrors and focused on a ZnTe detector. A probe beam with 35 fs pulses is also focused on the ZnTe detector. The electric-optical sampling signal in frequency domain is
\begin{equation}
\frac{\Delta I(\Omega)}{I}=\frac{\omega n^3 r_{41} L t_p}{2 c} E_{THz}(\Omega).
\end{equation}
Since ZnTe and CoSi are measured in the same setup, they should have the same collection efficiency for a specific THz frequency, thus
\begin{equation}
\frac{( \frac{\Delta I(\Omega)}{I})^{CoSi}}{ (\frac{\Delta I(\Omega)}{I})^{ZnTe}}=\frac{E_{eff}^{CoSi}(\Omega)}{E_{eff}^{ZnTe}(\Omega)}.
\end{equation}

Combining the equations above,
\begin{equation}
\eta(\Omega)=\frac{( \frac{\Delta I(\Omega)}{I})^{CoSi}}{ (\frac{\Delta I(\Omega)}{I})^{ZnTe}}\frac{n^{CoSi}(\Omega)}{n^{ZnTe}(\Omega)}\frac{T_p^{ZnTe}}{T_p^{CoSi}}\frac{l^{ZnTe}}{l^{CoSi}} \frac{\gamma^{ZnTe}}{\gamma^{CoSi}}\Omega\epsilon_0\chi_{41}
\end{equation}

\subsection{Derivation of effective emission depth}
Incident ultrafast light pulses consist of a distribution of frequency around the center photon energy. Generating current in CoSi (and polarization in the case of ZnTe) in the THz frequency involves with difference frequency generation (DFG) within the pulse frequency.

When two electric fields ($E_a$ and $E_b$) with frequency $\omega_{a}=\omega+\Omega/2$ and $\omega_{b}=\omega-\Omega/2$ are incident on the sample surface, the electric fields inside the sample at a depth of $d$ is $E_a=E_{a0}e^{ik_ad/\cos{(\theta_{in})}-i (\omega+\Omega/2) t}$ and $E_b=E_{b0}e^{ik_bd/\cos{(\theta_{in})}-i (\omega-\Omega/2) t}$. The generated current with THz frequency $\Omega$ at the depth of $d$ is
\begin{equation}
j_\Omega (t)=\sigma(\Omega) E_{a0}E_{b0}e^{i(k_a-k_b^*)d/\cos{(\theta_{in})}}e^{-i\Omega t}.
\end{equation}

The THz wave generated by the current in depth $d$ on the detector is
\begin{equation}
E_{\Omega} (t)=\Gamma({\Omega})\sigma(\Omega) E_{a0}E_{b0}e^{i(k_a-k_b^*)d/\cos{(\theta_{in})}}e^{-ik_\Omega d/\cos{(\theta_{out})}-i\Omega t},
\end{equation}
where $\Gamma(\Omega)$ is the conversion efficiency from THz current in the sample to THz wave on the detector at frequency $\Omega$.
After integrating over propagation direction of output THz wave, one gets
\begin{equation}
E_{\Omega}(t)=\frac{\Gamma({\Omega})\sigma(\Omega) E_{a0}E_{b0}e^{i(k_a-k_b^*)e^{-i\Omega t}}}{i(k_a-k_b^*)d\frac{\cos{\theta_{in}}}{\cos{\theta_{out}}}-ik_\Omega d}
\end{equation}

So the effective emission depth is
\begin{equation}
l_{\Omega}=\left |\frac{1}{i(k_a-k_b^*)d\frac{\cos{\theta_{in}}}{\cos{\theta_{out}}}-ik_\Omega d} \right |
\end{equation}

In the case of ZnTe, one can assume $k_a$ and $k_b$ to be purely real so
\begin{equation}
l^{ZnTe}(\Omega)=\frac{c}{(n_g(\omega)\frac{\cos(\theta_{out})}{\cos(\theta_{in})}+n^{ZnTe}(\Omega))\Omega},
\end{equation}
where $n_g(\omega)$ is the group index of ZnTe at the frequency $\omega$.

In the case of CoSi, $k_a$ and $k_b$ have the imaginary part $\Im(n^{CoSi}(\omega))\omega/c$ and is much larger than $(n_g+k_\Omega)\Omega/c$, so
\begin{equation}
l^{CoSi}(\Omega)=\frac{c}{2\Im(n^{CoSi}(\omega))\frac{\cos(\theta_{out})}{\cos(\theta_{in})}\omega}.
\end{equation}
{The excitation energy dependence of $l^{CoSi}(\Omega)$ is shown in Fig. \ref{parameters}d. Note in the figure we ignore its dependence on $\theta_{out}$ by the approxiamtion $\theta_{out}\ll1$. See details in the Sec. \ref{secparameters} }

\subsection{{Parameters based on optical  conductivity in CoSi}}
\label{secparameters}
During the above conversion process, many parameters relied on the optical conductivity in both THz and excitation energy range are needed. Our linear conductivity measurement in CoSi is summarized and discussed detailly in Ref. \cite{xuarXiv2020}. In this paper, we reproduce the data measured at room temperature at both the excitation energy range (Fig. \ref{parameters}a) and the THz radiation range (Fig. \ref{parameters}b). For excitation energy range, we calculate the refractive index (Fig. \ref{parameters}c), penetration depth (Fig. \ref{parameters}d), Fresnel correction coefficient (Fig. \ref{parameters}e) and refracted angle $\theta_{in}$ (Fig. \ref{parameters}f) as a function of excitation energy. As for the THz range, the lowest energy we get from the linear conductivity measurement is 1.6 THz, while the THz wave we detect in the THz emission experiment is 0.2-2 THz. To get the linear response below 1.6 THz, we fit the measured data from 1.6-6.0 THz by Drude formula. The zero-frequency extrapolation agrees with the DC transport conductivity. The best fit is shown in Fig. \ref{parameters}a with the broadening $\Gamma=\hbar/\tau$=14 meV and transport conductivity $\sigma_0=6.1\times 10^3\ \Omega^{-1}$cm$^{-1}$. We then derive the refractive index (Fig. \ref{parameters}g), the angle of in-phase radiation direction $\theta_{out}$ (Fig. \ref{parameters}h) and the Fresnel coefficient (Fig. \ref{parameters}i) in the THz range.

\subsection{{THz time traces of optical rectification in ZnTe and CPGE in CoSi at different excitation photon energy}}

{Here we show the raw data of THz waveform in the time domain of optical rectification in ZnTe and CPGE contribution of CoSi at 0.200 -1.050  eV in Fig. \ref{rawdata}. All the data are normalized by the incident power. ZnTe is excited by  horizontally polarized (xz-plane) incident laser pulses, and the in-plane component THz wave is collected. The CoSi CPGE data shown is one half of the difference between the in-plane THz wave of the left-handed and right-handed incident laser excitation pulses. The laser pulse energy range of 0.200-0.450 eV is generated by difference frequency generation (DFG), while the range of 0.475-1.05 eV is generated by optical parametric amplifier (OPA). In the DFG range, a germanium focusing lens and a MgF$_2$ achromatic quarter-wave plate are used, and in the OPA regime, a BaF$_2$ focusing lens and a quartz-MgF$_2$ achromatic quarter-wave plate are used. (See more details in Methods). A broadening of the waveform and a decreasing of the peak signal are observed at the low-energy edge of both DFG and OPA, which is caused by a larger laser pulse width from the total dispersion of the quarter-wave plate, the focusing lens and the linear polarizer. In general, the ZnTe shows an energy-independent signal as expected, which serves as a good benchmarking material. For the CoSi CPGE signal, the raw signal increases dramatically from 1.050 eV and peaks at around 0.400 eV.}

\begin{figure}

\centering
\includegraphics[width=\textwidth]{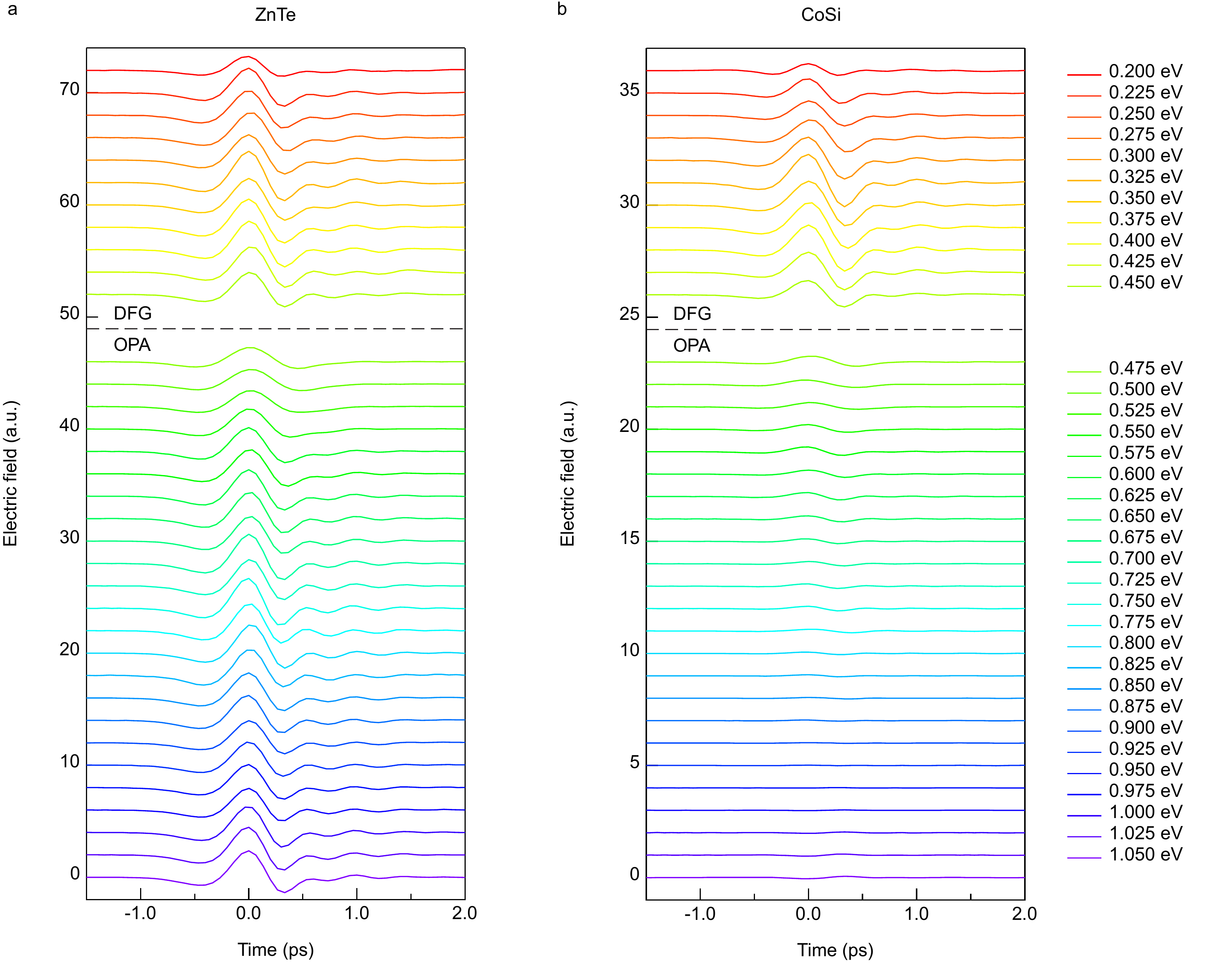}
\caption{{Raw data of THz time traces of optical rectification in ZnTe and CPGE in CoSi at different excitation photon energy. The data are normalized by the incident power.}}
\label{rawdata}
\end{figure}

\section{Additional symmetry-related measurement}
\subsection{{CPGE signal at different sample azimuth angles at 0.35 eV}}

{Besides the measurement performed at 0.50 eV shown in the main text Fig. 2g, CPGE signals at different sample azimuth angles are also measured at 0.35 eV, which is close to the CPGE peak. The measured THz time traces in different azimuth angles are shown in Fig. \ref{0.35eV}. All of the curves overlap well, which is consistent with the symmetry analysis.}

\begin{figure}

\centering
\includegraphics[width=0.6\textwidth]{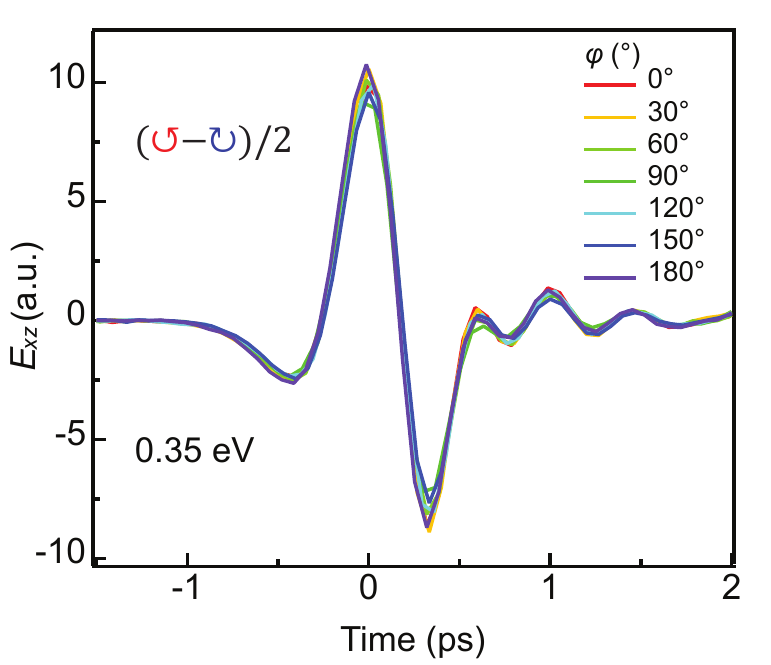}
\caption{{CPGE THz time traces at different sample azimuth angles $\phi$ at the incident photon energy of 0.35 eV.}}
\label{0.35eV}

\end{figure}

\subsection{THz peak signal at 0.375 eV as a function of quarter-wave plate angle}

\begin{figure}
\centering
\includegraphics[width=\textwidth]{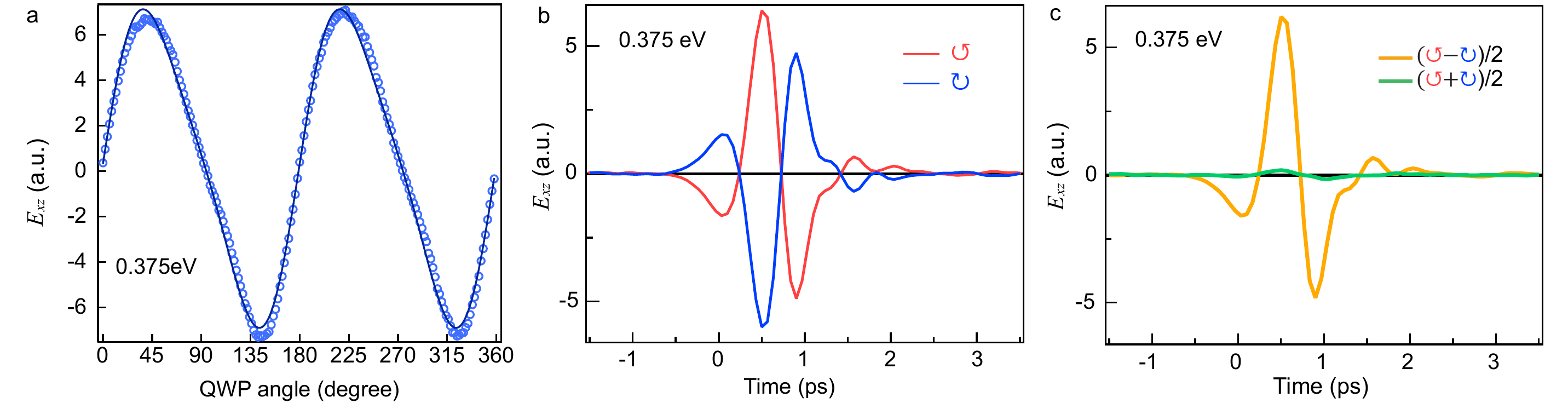}
\caption{a, CPGE THz peak amplitude in CoSi as a function of the angle of the quarter wave plate with incident photon energy of 0.375 eV. {b, THz wave form of the in-plane component under left-handed and right-handed incident laser pulses at 0.375 eV. c, CPGE and LPGE THz wave form of the in-plane component at 0.375 eV. The black line is at zero.} }
\label{FigS1}
\end{figure}

The measured $E_{xz}$ as a function of the quarter-wave plate angle is shown in Fig.\ref{FigS1}a. {The in-plane THz wave form under left-handed and right-handed laser pulses are shown in Fig.\ref{FigS1}b. The two curves are almost identical to each other with the opposite sign. The extracted CPGE and LPGE components are shown in Fig.\ref{FigS1}c.} From these two plots, one can see that CPGE much larger than LPGE at 0.375 eV. The black curve in Fig.\ref{FigS1} is the fit by $E_{xz}(\theta)=A\sin(2\theta)+B_2\sin(4\theta)+C_2\cos(4\theta)+D_2$. Note that $E_y(\theta)=B_1\sin(4\theta)+C_1\cos(4\theta)+D_1$. At 0.5 eV, fitting $E_{xz}$ and $E_y$ leads to $A_2 \colon B_2 \colon C_2 \colon D_2=155.25 \colon 78.09 \colon 7.42 \colon 20.84$ and $A_1 \colon B_1 \colon C_1 \colon D_1=0.51 \colon -5.86 \colon 77.33 \colon 117.62$. At 0.375 eV, fitting $E_{xz}$ gives rise to $A_2 \colon B_2 \colon C_2 \colon D_2=380.73 \colon 38.32 \colon 2.98 \colon 8.37$.

\section{Comparison with Quantized CPGE\label{sec:comp}}

The CPGE current is also called the injection current, of which the generation rate is proportional to incident power. The predicted quantized CPGE current satisfies
\begin{equation}
\frac{\partial{j}}{\partial{t}}=\beta_{xx}E_0^2(t).
\end{equation}
where $\beta_{xx}=i\beta_0/3=i\frac{\pi e^3}{3 h^2}$ in the quantization regime. If we assume the hot carrier lifetime in the sample to be $\tau$, then the Drude model for the current in the frequency domain is
\begin{equation}
i\Omega j(\Omega)=\beta_{xx}\mathcal{F}(E_0^2(t))-\frac{j(\Omega)}{\tau}
\end{equation}
which yields
\begin{equation}
j(\Omega)=\frac{\beta_{xx}}{i\Omega+1/\tau}\mathcal{F}(E_0^2(t))\equiv \eta \mathcal{F}(E_0^2(t)) ,
\end{equation}
where $\mathcal{F}(E_0^2(t))$ is the Fourier transform of $E_0^2(t)$, and where we have defined $\eta=\frac{\beta_{xx}}{i\Omega+1/\tau}$. As shown in Fig.~\ref{FigS2}a, we did not observe an obvious THz frequency dependency of $\eta$, which is consistent with the fact that $\tau$ is much shorter than the laser pulse width, justifying the approximation $\eta\approx \beta_{xx}\tau$.

Since the measured peak of $\beta_{xx}$ is 1.1C, in units of the quantization constant $\beta_0=\frac{\pi e^3}{h^2}$, and the measured $\beta \tau$ is around $550\mu A/V^2$, we have
\begin{align}
\beta \tau &=\beta \frac{\hbar}{\Gamma}=1.1 \frac{\pi e^3}{h^2} \frac{\hbar}{\Gamma}\\
&=1.1\frac{\pi e^3}{4\pi^2 \hbar eV}\frac{eV}{\Gamma}\\
&=21 \frac{\mu A}{V^2} \frac{eV}{\Gamma},
\end{align}
Therefore, we determine the broadening factor $\Gamma$ to be 38 meV, which corresponds to a relaxation time of $\tau=\frac{\hbar}{\Gamma}=17 fs$. As shown in Fig.~\ref{FigS2}b, when the chemical potential is at -37 meV, calculations with broadening of 5 meV and 38 meV could reproduce the profile (width) of experimental data, but apparently, the calculation of 5 meV broadening would give rise to a peak which is 7-8 times too high considering the extended relaxation time. Therefore, as stated in the main text, the broadening of 38 meV is chosen to constrain the width and the peak of the measured CPGE spectrum. In general, the hot-electron lifetime, and hence $\Gamma$, is energy-dependent. However, as shown in Fig. 3a in the main text, the assumption of an energy-independent broadening of 38 meV results in a calculated CPGE that matches well the experimental data between 0.25 eV and 0.7 eV. Calculations of different chemical potential with a 38 meV broadening are shown in Fig.~\ref{FigS2}c. The curve with $E_f=-37$ meV matches the photon-energy dependence well, and therefore we interpret this value as the Fermi level of our CoSi sample.

\begin{figure}
\centering
\includegraphics[width=\textwidth]{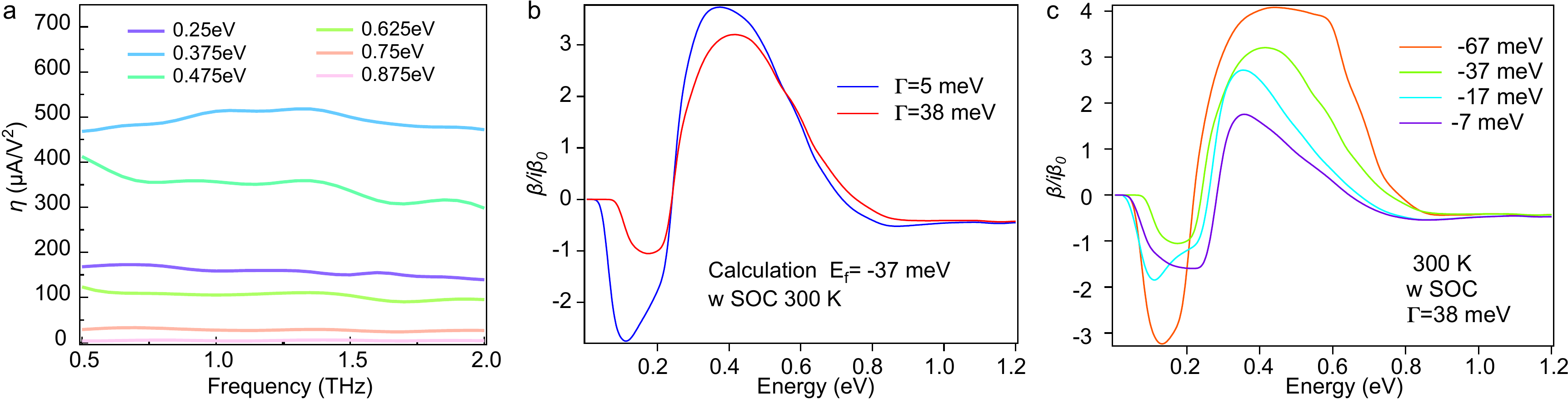}
\caption{a, CPGE current amplitude in the frequency domain at different incident photon energy in CoSi. b, CPGE calculation at room temperature with SOC of two different broadening of 38 meV and 5 meV. c, CPGE calculation at room temperature with SOC of broadening of 38 meV with different chemical potential.}

\label{FigS2}
\end{figure}

\section{Possible CPGE quantization from the DFT\label{app:SOC}}
Fig.\ref{FigS3}a shows the band structure of CoSi with spin-orbit coupling (SOC). The threefold nodes split into a four-fold spin 3/2 node and a Weyl node with $\sim$20 meV separation, much smaller than the 114 meV splitting in RhSi, which indicates that SOC is relatively small in CoSi. Consequently, we use 20 meV as an estimate for the energy scale of SOC in this compound. Fig.\ref{FigS3}b shows the CPGE current calculation at 0 K with different chemical potential with a broadening of 5 meV without SOC. It shows a wide quantization plateau from the $\Gamma$ point when the Fermi level is above the threefold node, other than the narrow quantization region at $E_f=-37$ meV as shown in the main text.

We further study the effect of changing the hot electron lifetime $\tau$ on the CPGE spectrum. Fig.\ref{FigS3}c shows the CPGE current calculation at 0 K with the chemical potential of $-37$ meV but with different broadening factors $\hbar/\tau$. The quantized plateau at $\sim$ 100 meV disappears when the broadening reaches 10 meV. Moreover, Fig.\ref{FigS3}d shows that with the broadening of 40 meV, the dip around 0.2 eV is not quantized even when the chemical potential is above the multifold nodes at $\Gamma$. Therefore, longer hot-electron lifetime or electron doping is essential for the observation of quantized CPGE in CoSi. Another possibility discussed in the main text is the hole doping to $E_f=-67$ meV, which turns on the transitions for the multifold nodes at the R point. The resulting wide quantization plateau around $0.4$ eV is quite robust even up to room temperature. Its detection would not require a low-frequency and low-temperature measurement.

Since large SOC would remove the CPGE quantization, we also compared the results with/without SOC. Fig.\ref{FigS3}e-f shows the calculation of CPGE current at two different chemical potentials with and without SOC. One can see that the difference is quite small, especially in the experimental measurement regime of 0.2 eV - 1.1 eV, which proves CoSi is an ideal platform for the realization of quantized CPGE. The small differences between calculations performed with and without SOC support the validity of the spinless $k \cdot p$ model used in the main text, and the four-band tight-binding model in Section~\ref{sec:TB} used to reproduce the dip-peak structure of the ab-initio CPGE spectrum.

\begin{figure*}
\centering
\includegraphics[width=\textwidth]{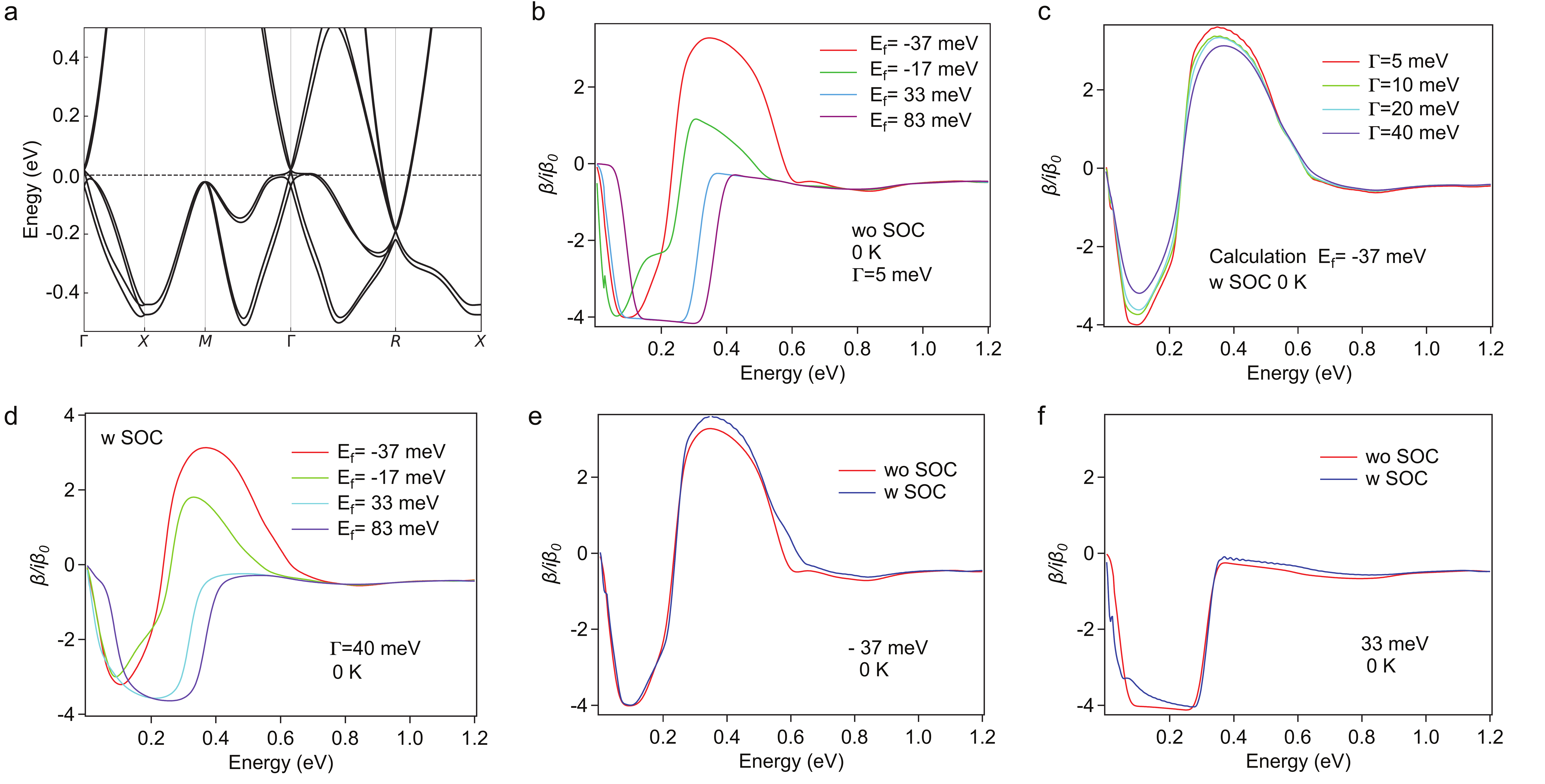}

\caption{ a, Band structure of CoSi with spin-orbit coupling. The dashed line is at $E_f=-37$ meV. Note that the DFT gives $E_f=-20$ meV. b, Calculated CPGE current at T=0 K with $E_f=-37$ meV, $E_f=-17$ meV, $E_f=33$ meV and $E_f=83$ meV with spin-orbit coupling. c, Calculated CPGE current at T=0 K with $E_f=-30$ meV with spin-orbit coupling and with different broadening. d, Calculated CPGE current at T=0 K with a broadening of 40 meV with spin-orbit coupling and with different chemical potential. Calculated CPGE current at T=0 K with 5 meV broadening with and without spin-orbit coupling at e, $E_f=-37$ meV and f, $E_f=33$ meV.}
\label{FigS3}
\end{figure*}

\section{CPGE from a tight-binding model for CoSi\label{sec:TB}}

\begin{figure}
\centering
\includegraphics[width=\linewidth]{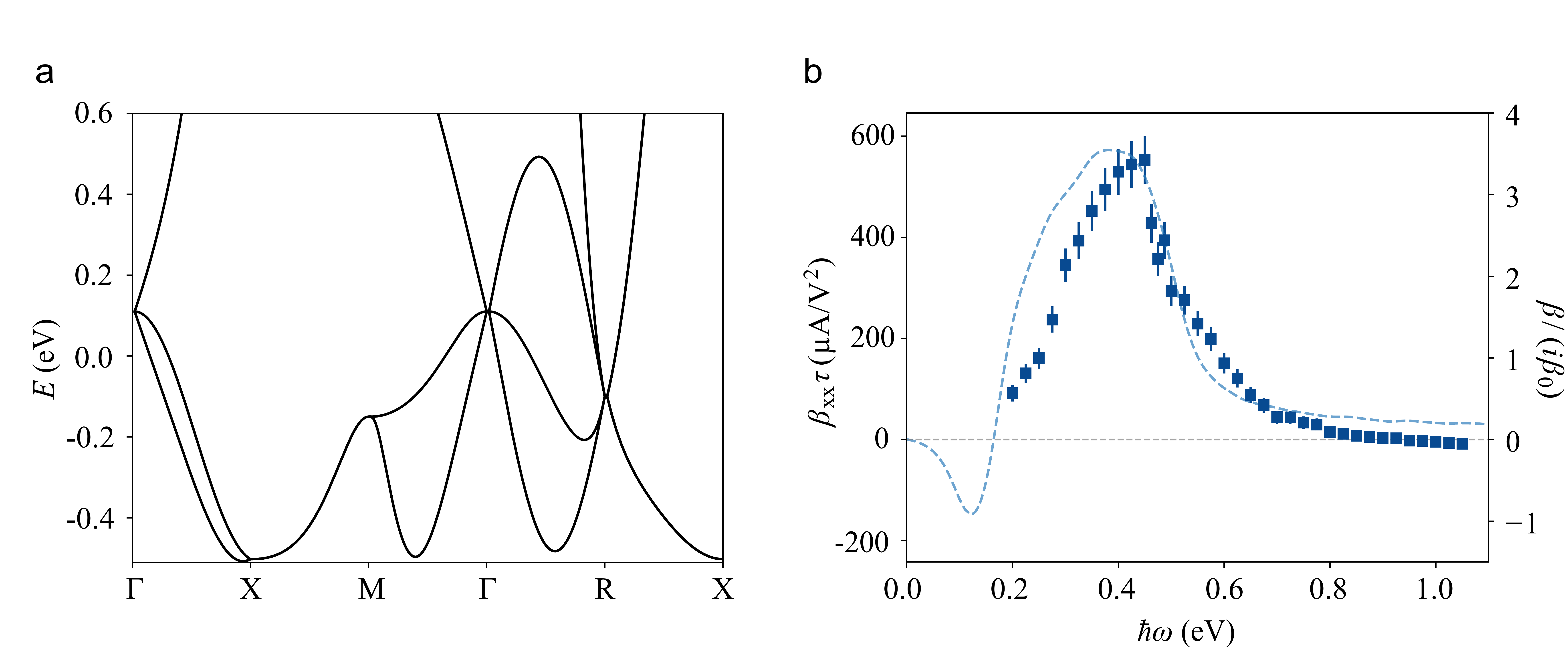}
\caption{a, Band structure obtained with the tight-binding model for space group 198 with parameters $v_1=1.29$, $v_p = 0.55$, and $v_2=0.25$ eV. $E=0$ marks the position of the chemical potential, $110$ meV below the threefold node at $\Gamma$. b, Corresponding CPGE obtained using Eq.~(2) in the main text with a broadening of $\Gamma=40$meV (dashed) compared to the experimental data (squares).}
\label{fig:CPGE_TB}
\end{figure}

In this section we calculate the CPGE using a tight-binding model that captures all symmetries of space group 198. This model was previously used to study RhSi in the same space group~\cite{changPRL2017,flicker_chiral_2018,SanchezMartinez:2019he}, and CoSi~\cite{xuarXiv2020}. Neglecting spin-orbit coupling (see Section.~\ref{app:SOC}), the tight-binding model is defined by three material-dependent parameters, $v_1$, $v_p$, and $v_2$, which we take as $v_1=1.29$, $v_p = 0.55$, and $v_2=0.25$ eV for CoSi~\cite{xuarXiv2020}. To this model we add a constant energy shift of $0.65$ eV, which amounts to choosing the chemical potential to be $110$ meV below the node at $\Gamma$. This chemical potential is chosen to reach the best agreement with the experimental data. The resulting bands are shown in Fig.~\ref{fig:CPGE_TB}a. Compared to DFT these bands capture well the linearly dispersing multifold bands and the separation of the multifold nodes in energy and momentum space. However, this model overestimates the bandwidth of the multifold bands, and the quadratic corrections to the flat band at $\Gamma$ compared to DFT.

The calculation of the CPGE follows Ref.~\cite{flicker_chiral_2018} and uses Eq.~(2) in the main text. To incorporate the effect of disorder, we add a phenomenological broadening of the Dirac delta function $\delta(x)\to \frac{1}{\pi}\frac{\Gamma}{x^2+\Gamma^2}$, where $\Gamma=\hbar/\tau$. To compare with experiment, we plot $\beta
_{xx}\tau$ as discussed in the Methods and Section~\ref{sec:comp}. The result with $\Gamma=40$ meV ($\tau\approx 16$ fs) and $T=0$ is shown in Fig.~\ref{fig:CPGE_TB}b. The curve shows the peak-dip feature discussed in the main text and in Section~\ref{sec:kdotp}.

\section{\label{sec:kdotp}Spinless $k \cdot p$ model with quadratic corrections}

As argued in the main text, one can gain more insight into the main features of the experiment using an effective $k \cdot p$ model neglecting the effect of SOC, where the band structure of CoSi has a threefold crossing at $\Gamma$ and a double Weyl crossing at $R$. The double Weyl crossing at $R$ has linear dispersion for all energies of interest, so the relevant model is that of Ref. \cite{flicker_chiral_2018}, which is explicitly
\begin{align}
H_R = \left(\begin{array}{cc} v_R \vec \sigma \cdot \vec k & 0 \\ 0 & v_F \vec \sigma \cdot \vec k \end{array} \right)
\end{align}
The spinless threefold model at $\Gamma$ has also been presented in Ref. \cite{flicker_chiral_2018} but only to linear order in momentum. For this work, we need to extend it up to quadratic order, which we do next.

CoSi has space group 198, and for effective models near $\Gamma$ we only need the irreducible representations (irreps) of the point group, which is $T$. However, it will be useful to first derive the model for the point group of higher symmetry $O$, and then find the terms that break the symmetry down to $T$. This will clarify the relative importance of the different quadratic corrections.

The point group $O$ can be generated by $C_3$ rotations around (111), $C_2$ rotations around (100) and $C'_2$ rotations around (110), and it has 5 irreps transforming as $A_1 \sim x^2 + y^2 + z^2$, $A_2 \sim xyz$, $E \sim (x^2-y^2, (2z^2-x^2-y^2)/\sqrt{3})$, $T_1 \sim (x,y,z) $ and $T_2 \sim (yz,zx,xy)$. Point group $T$ is obtained by breaking $C'_2$, so it is generated by the $C_3$ and $C_2$ operations only, and has three irreps, $A$, $E$, $T$. The subscript in $A_{1,2}$ and $T_{1,2}$ distinguishes the transformation under $C'_2$ in $O$, and it is absent in $T$ where these pairs of irreps become the same.

The three basis states for the threefold crossing at $\Gamma$ in $O$ form a $T_1$ irrep. Operators acting in the subspace of this three basis states can be chosen as the Gell-Mann matrices $\lambda_\alpha$ with $\alpha=1,\cdots,8$ (and $i=0$ the identity), which are given by
\begin{align}
\lambda_1 &= \left( \begin{matrix} 0 & 1 & 0 \\ 1 & 0 & 0 \\ 0 & 0 & 0 \end{matrix} \right) \qquad &
\lambda_2 &= \left( \begin{matrix} 0 & -i & 0 \\ i & 0 & 0 \\ 0 & 0 & 0 \end{matrix} \right) \qquad &
\lambda_3 &= \left( \begin{matrix} 1 & 0 & 0 \\ 0 & -1 & 0 \\ 0 & 0 & 0 \end{matrix} \right) \qquad &
\lambda_4 &= \left( \begin{matrix} 0 & 0 & 1 \\ 0 & 0 & 0 \\ 1 & 0 & 0 \end{matrix} \right) \\
\lambda_5 &= \left( \begin{matrix} 0 & 0 & -i \\ 0 & 0 & 0 \\ i & 0 & 0 \end{matrix} \right) \qquad &
\lambda_6 &= \left( \begin{matrix} 0 & 0 & 0 \\ 0 & 0 & 1 \\ 0 & 1 & 0 \end{matrix} \right) \qquad &
\lambda_7 &= \left( \begin{matrix} 0 & 0 & 0 \\ 0 & 0 & -i \\ 0 & i & 0 \end{matrix} \right) \qquad &
\lambda_8 &= \dfrac{1}{\sqrt{3}}\left( \begin{matrix} 1 & 0 & 0 \\ 0 & 1 & 0 \\ 0 & 0 & -2 \end{matrix} \right)
\end{align}
They transform as bilinears of the basis states so they must form irreps $T_1 \otimes T_1 = A_1 + E + T_1 + T_2$. Taking the standard representation of the rotation operators as
\begin{align}
C_{3,(111)} = \left(\begin{array}{ccc} 0 & 0 & 1 \\ 1 & 0 & 0 \\ 0 & 1 & 0 \end{array} \right) ,
&& C_{2x} = \left(\begin{array}{ccc} -1 & 0 & 0 \\ 0 & -1 & 0 \\ 0 & 0 & 1 \end{array} \right) ,
&& C'_{2 (110)} = \left(\begin{array}{ccc} -1 & 0 & 0 \\ 0 & 0 & 1 \\ 0 & 1 & 0 \end{array} \right) ,
\end{align}
it can be checked that the following combinations of Gell-Matrices transform as the irreps of point group $O$:
\begin{align}
T_1 &= (-\lambda_2,\lambda_5,-\lambda_7) , \\
T_2 &= (\lambda_1,\lambda_4,\lambda_6) , \\
E &= (-\tfrac{1}{2}\lambda_3 + \tfrac{\sqrt{3}}{2} \lambda_8,-\tfrac{\sqrt{3}}{2} \lambda_3 -\tfrac{1}{2}\lambda_8) ,
\end{align}
while $\lambda_0$ trivially transforms as $A_1$. Time reversal symmetry is implemented as complex conjugation, so $\mathcal{T}^{-1} T_1 \mathcal{T} = -T_1$ is odd while $\mathcal{T}^{-1} T_2 \mathcal{T} = T_2$ and $\mathcal{T}^{-1} E \mathcal{T} = E$ are even. The matrices in $T_1$ are also the spin-1 matrices $\vec S$ used in the main text. The effective Hamiltonian can now be built making scalar combinations of the Gell-Mann matrices with momentum irreps, which up to second order are
\begin{align}
K_{A1} &= k_x^2 + k_y^2 + k_z^2 , \\
K_{T1} &= (k_x,k_y,k_z) , \\
K_{T2} &= (k_y k_z, k_x k_z, k_x k_y) , \\
K_E &= (k_x^2-k_y^2, (2k_z^2-k_x^2-k_y^2)/\sqrt{3}) ,
\end{align}
which allows four terms in the Hamiltonian preserving time-reversal symmetry
\begin{align}
H_O = \left(\begin{array}{ccc}
a k^2 + \dfrac{2c}{3}(k^2 -3k_z^2) & i v k_x + b k_y k_z & -i v k_y + b k_x k_z \\
-i v k_x + b k_y k_z & a k^2 + \dfrac{2c}{3}(k^2 -3k_y^2) & i v k_z + b k_x k_y \\
i v k_y + b k_x k_z& -i v k_z + b k_x k_y & a k^2 + \dfrac{2c}{3}(k^2 -3k_x^2)
\end{array}\right) ,
\end{align}
where $k=\sqrt{k_{x}^{2}+k_{y}^{2}+k_{z}^{2}}$. If we now consider the physical point group $T$, a single extra term is allowed because there is
a new momentum irrep $K'_E = (-(2k_z^2-k_x^2-k_y^2)/\sqrt{3},k_x^2-k_y^2)$, which leads to
\begin{align}
H_T = H_O + \dfrac{2d}{\sqrt{3}} \left(\begin{array}{ccc}
k_y^2-k_x^2 &0& 0\\
0&k_x^2 - k_z^2& 0\\
0&0&k_z^2 - k_y^2
\end{array}\right) .
\end{align}

To obtain the values of this coefficients, we expand the energies of the three bands to lowest order in momentum. The exact energy of any three band Hamiltonian written in terms of the Gell-Mann matrices $H=\lambda_0 h_0 + h_{\alpha}\lambda_{\alpha}$ is given by
\begin{equation}
E_{n} = h_0 + 2\sqrt{\dfrac{h_{2}}{3}} \cos \left[ \dfrac{1}{3}\arccos\left( \dfrac{h_{3}}{h_{2}}\sqrt{\dfrac{3}{h_{2}}} \right) + \dfrac{2\pi n}{3} \right] \ ,
\end{equation}
where $\alpha=1,\dots,8$, $h_{2}=h_{\alpha}h_{\alpha}$, $h_{3}=d_{\alpha\beta\gamma}h_{\alpha}h_{\beta}h_{\gamma}$, and $d_{\alpha\beta\gamma}$ are the SU(3) symmetric structure constants. Expanding the energies up to second order in $k$ we find
\begin{align}
E_{1} = & \ -v k +a k^2 - \left( b + \dfrac{2c}{3} \right) f_{1}(\vec{k}) k^{2} + \dfrac{2c}{3} f_{2}(\vec{k}) k^{2} %- \dfrac{1}{2a} \left( (b+2c)^{2}f_{1}(\vec{k}) + \dfrac{4d^{2}}{3} \right) \left( f_{2}(\vec{k}) - f_{1}(\vec{k} ) \right) k^3
, \\
E_{2} = & \ a k^2 + 2\left( b + \dfrac{2c}{3} \right) f_{1}(\vec{k}) k^{2} - \dfrac{4c}{3} f_{2}(\vec{k}) k^2 , \\
E_{3} = & \ v k +a k^2 - \left( b + \dfrac{2c}{3} \right) f_{1}(\vec{k}) k^2 + \dfrac{2c}{3} f_{2}(\vec{k}) k^2 %+ \dfrac{1}{2a} \left( (b+2c)^{2}f_{1}(\vec{k}) + \dfrac{4d^{2}}{3} \right) \left( f_{2}(\vec{k}) - f_{1}(\vec{k} ) \right) k^3
,
\end{align}
where $k=\sqrt{k_{x}^{2}+k_{y}^{2}+k_{z}^{2}}$, $f_{1}(\vec{k})=\left( k_{x}^{2}k_{y}^{2}+k_{y}^{2}k_{z}^{2}+k_{z}^{2}k_{x}^{2} \right) /k^{4}$ and $f_{2}(\vec{k})=\left( k_{x}^{4}+k_{y}^{4}+k_{z}^{4} \right) /k^{4}$. We observe that there are no terms proportional to $d$ in this expansion to second order, which means $d$ cannot be obtained by fitting the bands alone. However, this does not mean that we can set $d$ to zero, as the matrix elements of the Hamiltonian still depend on $d$. In this model, to fit $d$ from ab-initio input would require an ab-initio calculation of the low energy Berry curvature or some other matrix element that is sensitive to $d$, as it was done in \cite{cook2017design}. For our purposes, we leave $d$ as a free parameter, and check that it has a negligible influence on the CPGE. In the main text, for simplicity we present the model with $d=0$. Inclusion of this term would have only been important to describe effects that are forbidden under $O$ but finite under $T$, like the linear photo-galvanic effect (LPGE) or second harmonic generation (SHG).

From ab-initio computations, we obtain the value of $v$, $a$, $b$ and $c$ by fitting the energy bands near the $\Gamma$ point. The results to order $k^{2}$ are shown in Figs.~\ref{fig:GR_BestFitk3}.

\begin{figure}[!t]
\centering
\includegraphics[width=\textwidth]{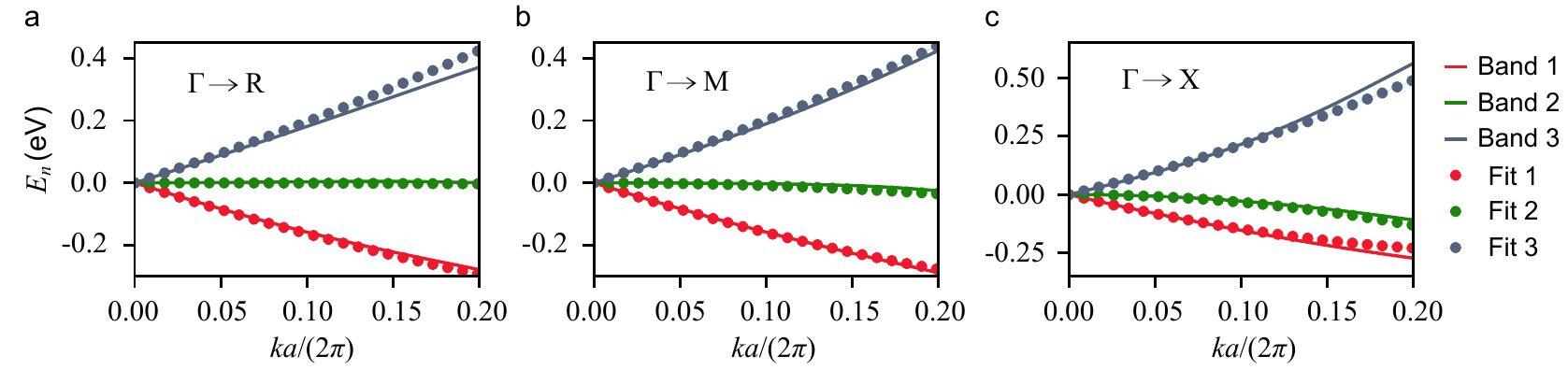}

\caption{Fit of the $k\cdot p$ model parameters from the {\sl ab-initio} energy bands, with the result $(v,a,b,c) = (1.79,1.07,-1.72,3.26)$.}
\label{fig:GR_BestFitk3}
\end{figure}

We can now proceed to compute the CPGE using Eq. 2 of the main text. For the R point, since we assume a linear double Weyl model with no tilt due to time-reversal, the CPGE has been computed before \cite{flicker_chiral_2018}, and it is given by a sharp step function $\theta(\omega-\omega_0)$. The position of the step $\omega_0$ can be estimated, to a very good approximation, by the energy difference in the R-X direction where there is no splitting to any order. When $\mu$ crosses the node, we obtain $\omega_0 = 286$ meV, and for $\mu = - 10$ meV we obtain $\omega_0 = 267$ meV.

For the $\Gamma$ point calculation, quadratic corrections have two effects: first, they modify the energies inside the $\delta$ function and Fermi functions, modifying the allowed transitions and thus the JDOS. Second, they modify the matrix elements in the integrand, spoiling the perfect quantization of the linear model. In Fig.~\ref{fig:CPGE_various}b we calculate the CPGE for the $\Gamma$ point considering only the first effect. Because the matrix elements are that of the linear model, the transition $1\rightarrow 3$ is forbidden by angular momentum conservation. For the other two transitions $1\rightarrow 2$ and $2\rightarrow 3$, the effect of quadratic corrections is to modify the energies where they become active or inactive.
These critical frequencies are determined by the dispersion in specific directions where the energy differences between bands are minimal or maximal, for a given modulus of $\vec{k}$. For our ab-initio parameters and for sufficiently small negative chemical potential, the directions where energy differences are minimal or maximal are always $\Gamma-X$ and $\Gamma-R$, respectively (see Fig.~\ref{fig:CPGE_various}a). Since bands $1$ and $2$ are partially filled, there
is a frequency window $(\omega_{1}^{X},\omega_{1}^{R})$ from where the transition $1\rightarrow2$ activates to where it becomes maximal, and another frequency window $(\omega_{2}^{X},\omega_{2}^{R})$ where its contribution decreases to zero. On the contrary, band 3 is always empty, so the transition $2\rightarrow3$ activates in the frequency window $(\omega_{3}^{X},\omega_{3}^{R})$ and does not disappear at larger frequencies. These critical frequencies can be found analytically to first order in the chemical potential and read
\begin{align}
\omega_{1}^{X} = & -\mu \left( 1 + \dfrac{4c - 3a}{3v^{2}}\mu \right) \ , \\
\omega_{2}^{X} = & \dfrac{6c}{4c-3a}\mu + v\sqrt{\dfrac{-3\mu}{4c-3a}} \ , \\
\omega_{3}^{X} = & \dfrac{-6c}{4c-3a}\mu + v\sqrt{\dfrac{-3\mu}{4c-3a}} \ , \\
\omega_{1}^{R} = & -\mu \left( 1 - \dfrac{2b+3a}{3v^{2}} \mu \right) \ , \\
\omega_{2}^{R} = & \dfrac{3b}{2b+3a}\mu + v\sqrt{\dfrac{3\mu}{2b+3a}} \ , \\
\omega_{3}^{R} = & \dfrac{-3b}{2b+3a}\mu + v\sqrt{\dfrac{3\mu}{2b+3a}} \ .
\end{align}
In our CoSi effective model with the fitted ab-initio parameters, we find that $\mu=-40$ meV is well within the applicability of these equations. The transition $1\rightarrow 2$ starts to die out before the $2\rightarrow 3$ picks up, which leaves a dip in the CPGE.
Note that we have not included $\omega_{2}^{R}$ and $\omega_{3}^{R}$ in the main text because they take too large values due to the flatness of the intermediate band.
To compute the total CPGE we also need the contribution from the R point, which produces a sharp jump. When this sharp jump accidentally occurs in the middle of the dip contributed by $\Gamma$, we generically get a dip-peak structure as observed ab-initio (see Fig.~\ref{fig:CPGE_various}c).
For different model parameters, the critical frequencies might be determined by the dispersion in other directions different than $\Gamma-X$ and $\Gamma-R$, and can only be calculated numerically.

Including now the corrections to the integrand (see Fig.~\ref{fig:CPGE_various}d), we observe that they only lead to a smooth change that grows with frequency, without changing the curves qualitatively. Transitions from band 1 to 3 are now allowed, but we found them negligible for our parameter set and were not included in any plot. The origin of the dip can therefore be attributed to the change in JDOS induced by quadratic corrections. The dip remains approximately quantized as it originates from transitions $1\rightarrow 2$ with a closed manifold. In contrast, the peak is generically non-universal since it is composed from contributions of both the $\Gamma$ and $R$. If the frequency window where the $\Gamma$ contribution vanishes overlaps with the window where the $R$ point transitions contribute, the peak becomes universal.

Finally, we have also studied the effect of the parameter $d$, which was set to $d=0$ in Fig.~\ref{fig:CPGE_various}d. We have recomputed all the curves in this figure with $d = 1$ and checked that for $\omega < 0.4 $ eV, the curves deviate from those at $d=0$ by 3\% at most, and this difference reduces monotonically for lower energies. As anticipated, the effect of this parameter is negligible in CPGE.

\begin{figure}[!t]
\centering
\includegraphics[width=0.7\textwidth]{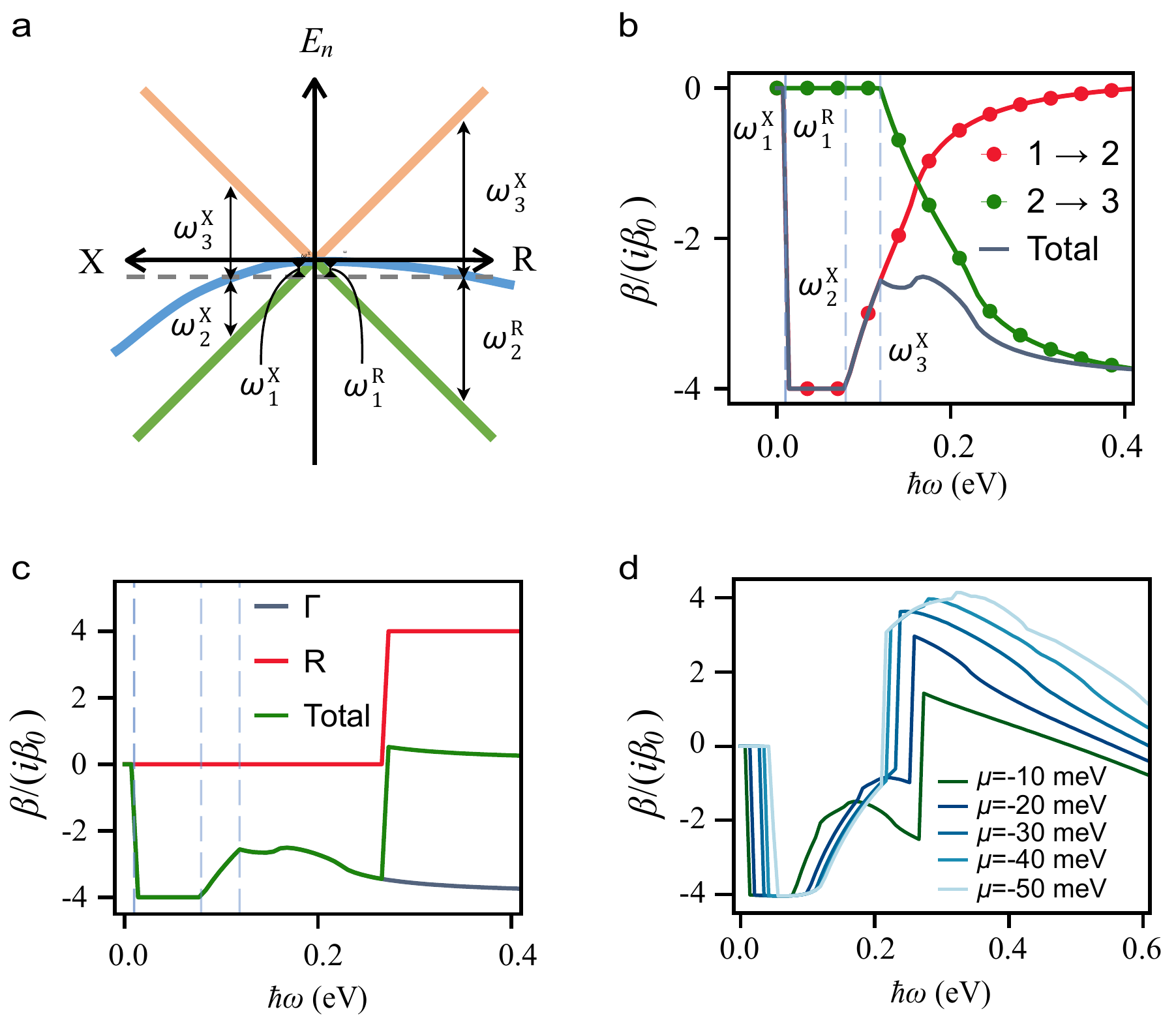}
\caption{Different CPGE results from the {\sl ab initio} fitting. a, Sketch of the critical frequencies for a negative chemical potential. b, $\Gamma$ point contribution coming from the first effect of quadratic corrections for $\mu=-10$ meV. The vertical lines show the critical frequencies corresponding to each band transition contribution. c, Total contribution from the $\Gamma$ and R points for $\mu=-10$ meV including the effect of quadratic corrections on the energies but not on the matrix elements, showing the characteristic shape of CPGE. d, CPGE for different chemical potentials, now including all the effects of quadratic corrections.}
\label{fig:CPGE_various}
\end{figure}

\end{widetext}

\end{document}